\documentclass[aoas,MSNbibl,nameyear,dvips]{arximspdf}
\usepackage{dcolumn,multirow}
\usepackage{graphicx}

\doi{10.1214/13-AOAS628} 
\volume{7}
\issue{2}
\pubyear{2013}
\firstpage{739}
\lastpage{762}
\setattribute{copyright}{owner}{In the Public Domain}

\makeatletter

\newcolumntype{d}[1]{D{.}{.}{#1}}
\newcommand{\balpha}{\bolds{\alpha}}
\newcommand{\boldeta}{\bolds{\rho}}
\newcommand{\btriangle}{\nabla}
\newcommand{\bX}{\mathbf{X}}
\newcommand{\bz}{\mathbf{z}}
\newcommand{\bs}{\mathbf{s}}
\newcommand{\bK}{\mathbf{K}}
\newcommand{\bu}{\mathbf{u}}
\newcommand{\bJ}{\mathbf{J}}
\newcommand{\ba}{\mathbf{a}}
\newcommand{\Frechet}{\mbox{Fr\'{e}chet}}

\newproclaim{assumption}{Assumption}
\newproclaim{definition}{Definition}

\newproclaim{example}{Example}
\newproclaim{remark}{Remark}

\makeatother

\begin{document}
\begin{frontmatter}

\title{Extreme value analysis for evaluating ozone control strategies\thanksref{T1}}
\runtitle{Extreme value analysis of ozone}
\thankstext{T1}{Supported in part by NIH Grant
R01-ES-014843-02 (Reich), EPA-STAR award R835228 (Reich and Cooley),
and the NSF (DMS-11-07046, Reich; DMS-09-05315, Cooley). The United
States Environmental Protection Agency through its Office of Research
and Development funded and managed the research described here (Foley
and Napelenok). It has been subjected to the Agency's administrative
review and approved for publication.}

\begin{aug}
\author[A]{\fnms{Brian} \snm{Reich}\corref{}\ead[label=e1]{reich@stat.ncsu.edu}},
\author[B]{\fnms{Daniel} \snm{Cooley}},
\author[C]{\fnms{Kristen} \snm{Foley}},
\author[C]{\fnms{Sergey}~\snm{Napelenok}}
\and
\author[D]{\fnms{Benjamin} \snm{Shaby}}
\runauthor{B. Reich et al.}
\affiliation{North Carolina State University,
Colorado State University, U.S. Environmental Protection Agency,
U.S. Environmental Protection Agency and University of California, Berkeley}
\address[A]{B. Reich\\
North Carolina State University\\
2501 Founders Drive, Box 8203\\
Raleigh, North Carolina 27695\\
USA\\
\printead{e1}} 
\address[B]{D. Cooley\\
Department of Statistics\\
Colorado State University\\
Ft. Collins, Colorado 80523-1877\hspace*{10pt}\\
USA}
\address[C]{K. Foley\\
S. Napelenok\\
National Exposure Research Laboratory\\
U.S. Environmental Protection Agency\\
Research Triangle Park, North Carolina 27711\\
USA}
\address[D]{B. Shaby\\
Department of Statistics\\
University of California, Berkeley\\
Berkeley, California 94720-3860\\
USA}
\end{aug}

\received{\smonth{8} \syear{2012}}
\revised{\smonth{12} \syear{2012}}

%
\begin{abstract}
Tropospheric ozone is one of six criteria pollutants regulated by the
US EPA, and has been linked to respiratory and cardiovascular endpoints
and adverse effects on vegetation and ecosystems. Regional
photochemical models have been developed to study the impacts of
emission reductions on ozone levels. The standard approach is to run
the deterministic model under new emission levels and attribute the
change in ozone concentration to the emission control strategy.
However, running the deterministic model requires substantial computing
time, and this approach does not provide a measure of uncertainty for
the change in ozone levels. Recently, a reduced form model (RFM) has
been proposed to approximate the complex model as a simple function of
a few relevant inputs. In this paper, we develop a new statistical
approach to make full use of the RFM to study the effects of various
control strategies on the probability and magnitude of extreme ozone
events. We fuse the model output with monitoring data to calibrate the
RFM by modeling the conditional distribution of monitoring data given
the RFM using a combination of flexible semiparametric quantile
regression for the center of the distribution where data are abundant
and a parametric extreme value distribution for the tail where data are
sparse. Selected parameters in the conditional distribution are allowed
to vary by the RFM value and the spatial location. Also, due to the
simplicity of the RFM, we are able to embed the RFM in our Bayesian
hierarchical framework to obtain a full posterior for the model input
parameters, and propagate this uncertainty to the estimation of the
effects of the control strategies. We use the new framework to evaluate
three potential control strategies, and find that reducing
mobile-source emissions has a larger impact than reducing point-source
emissions or a combination of several emission sources.
\end{abstract}

%

\end{frontmatter}

\section{Introduction}\label{sintro}

Due to advances in emissions control technology and regulatory action,
air quality has been improving over the last several decades in the
United States and Europe. However, areas exist where significant
populations are still exposed to elevated levels of tropospheric ozone
(O$_3$). Epidemiological and controlled human exposure studies have
shown an association between O$_3$ exposure and respiratory and
cardiovascular endpoints, particularly in sensitive populations [US
EPA (\citeyear{epa-2006})]. Furthermore, ozone has been linked to a
variety of adverse effects on vegetation and ecosystems [US EPA (\citeyear{epa-2006})].

Ozone, together with other compounds, is formed downwind of its two
main classes of precursors, volatile organic compounds (VOCs) and
nitrogen oxides (NO$_{\mathrm{x}} ={}$NO${}+{}$NO$_2$) in the presence of
sunlight. Due to the complexities of the formation processes of
secondary pollutants such as ozone and their dependence on various
physical and chemical parameters, as well as meteorological conditions,
ozone is a highly nonlinear function of its inputs and, thus, regional
three-dimensional Eulerian photochemical models have been developed to
track the precursor emissions, transport, and chemical transformations
of gases and particles in the troposphere. Such models are used to
study the formation and geographical distribution of ozone and also to
provide a test bed for possible control strategies. For example, we use
the Community Multiscale Air Quality (CMAQ) model [\citet{Byun2006}] to
study changes in ozone levels due to changes in mobile-source
NO$_{\mathrm{x}}$ emissions, point-source NO$_{\mathrm{x}}$ emissions,
other NO$_{\mathrm{x}}$ emissions, anthropogenic VOCs emissions, and
biogenic VOCs emissions. Due to the prevalence of ozone precursor
emissions from a wide range of sources, differing chemical reactivities
of the various specific chemical compounds that make up VOCs, and the
varying cost of different control technologies, the processes of
control strategy evaluation is itself complex.

Control strategy evaluation seeks to understand the potential response
in air quality levels from various targeted reductions of source
pollutants. By studying control strategies, we understand how resources
should be allocated to achieve the best results. Control strategies
must be studied via atmospheric chemistry models where it is possible
to change emissions scenarios. In the context of control strategy
evaluation, it is the response of modeled concentration that is desired
as a surrogate for the response for actual atmospheric pollutant
levels. Such response can be obtained most basically by performing two
modeling simulations: one representing current conditions; and one
representing conditions under a specific emissions control scenario.
The difference in predicted concentrations from the two simulations
could then be attributed to the control strategy. Currently, the
computational costs of running full regional photochemical models are
still nontrivial, making such evaluations potentially costly. To
address this issue, various methods have been applied to develop
reduced form models (RFMs) that represent pollutant concentrations for
a particular episode of interest as a simple function of usually only
the regulatory controllable parameters. For example, in the RFM, ozone
only may be a function of parameters for the various anthropogenic
sources NO$_{\mathrm{x}}$ and VOCs for a particular place and time. One
of the various methods for developing RFMs to date is through
calculating sensitivity coefficients of a target pollutant to emissions
of precursors from controllable sectors. These sensitivities are
subsequently used to make adjustments to pollutant concentrations
predicted by a base model simulation [\citet{digar-2011,Napelenok-2011}].

We propose a new approach for combining an RFM with monitored
point-level ozone data to study effects of various control strategies.
Our approach is geared toward accurately characterizing extreme ozone
events under different scenarios. Extreme ozone is a concern for health
effects modeling and regulation. For example, the current EPA
regulation of ozone is based on the fourth highest daily eight-hour
average ozone (i.e., the maximum eight-hour average ozone
concentrations for the day) of the year. Extreme value theory (EVT)
[see \citet{coles-2001a} for an overview] provides an
asymptotically-justified approach to modeling the tails of wide classes
of distributions. Our model employs a parametric form suggested by
extreme value theory in the tail, while utilizing a flexible quantile
regression approach to model the bulk of the distribution.

CMAQ output and monitor data are not directly comparable, as CMAQ
output is defined as a volume average within a three-dimensional grid
cell and we wish to make inference about ozone at point locations.
Thus, we build a downscaler, that is, a statistical model that links
gridded numerical model output to point-level observational data [see
\citet{berrocal-2010a} and references therein]. Most downscaling work
employs a Gaussian process framework [e.g., \citet{berrocal-2010a}] and
focuses on modeling the conditional mean of the observations given the
numerical model output. For example, \citet{foley-2012} use this
approach for ozone control-strategy evaluation. Some recent work has
sought to move beyond the parametric Gaussian paradigm. For example,
\citet{reich-2011} and \citet{zhou-2011a} propose separate models for
the quantile process of the model output and the quantile process of
the point-located data, which provides a calibration function to link
the two sources of data. \citet{mannshardt-2010a} propose an EVT-based
downscaling method for extreme precipitation, relating the return
levels (i.e., extreme quantiles, a climatological quantity) of the
model output to return levels at point-located sites. Unlike the
previous downscaling work for ozone which focused on calibration, our
focus is prediction. Specifically, once our downscaler is constructed
to relate monitor data to RFM output under observed emissions, we use
it to study the effect of alternative emission scenarios.

Other work which has appeared in the climate literature applies EVT to
numerical model output in order to produce maps summarizing the extreme
behavior of the studied variable. One approach [e.g., \citet
{kharin-2007a,wehner-2005a,wehner-2010a}] is to fit extreme value
distributions separately to the output at individual grid cells,
perhaps employing spatial smoothing procedures after the pointwise
fitting to produce the maps. Another approach [e.g., \citet
{cooley-2010a,schliep-2010a}] constructs hierarchical Bayesian models
that pool information across space by incorporating spatial random
effects. These works differ fundamentally in both aim and approach from
the present work. These previous studies aim to describe the tail of
the (unconditional) distribution and treat the numerical model output
as data, whereas we aim to model the conditional distribution of the
point-located measurements treating the model output as covariate information.


Our work is somewhat related to recent work which has sought to link
extreme behavior to large-scale climatological conditions. \citet
{sillmann-2011a} model the connection between extreme cold temperatures
and atmospheric blocking conditions as produced by both reanalysis
models and climate models.
\citet{maraun-2011a} link extreme precipitation to large-scale
airflow. Similar to the model we present in Section \ref{smodel},
\citet{sillmann-2011a} and \citet{maraun-2011a} both condition the
parameters of the extreme value distribution on the covariate
information. Both \citet{sillmann-2011a} and \citet{maraun-2011a} model
only the maximum value over a block of time, for example, monthly or
yearly maximum of daily value. In contrast, we use all observations to
both flexibly model the center of the distribution and model the upper
tail using EVT given any level of the covariate.

Our approach is novel for several reasons. First, the aim of our
analysis is to investigate control strategies. The RFM model is simple
enough to permit re-evaluation for a new set of input parameters at
negligible computing cost. As a result, we are able to embed the RFM
inside the Bayesian model to make inference on the input parameters
based on the resulting fit of the RFM to the monitor data, as in \citet
{Kennedy-2001}, \citet{Higdon-2004}, and \citet{foley-2012} for
Gaussian data. Therefore, unlike previous downscaler methods for
extremes [\citet{mannshardt-2010a,reich-2011,zhou-2011a}], we do not
estimate the densities of the model and monitor data separately. To
make full use of the RFM, we model the conditional distribution of the
monitor data given the RFM. Second, our model for the conditional
distribution of the monitor data given the RFM has attractive features.
EVT tells us that observations which exceed a high threshold are
well-approximated by the generalized Pareto (GPD) distribution.
Therefore, in contrast to most previous methods for downscaling
numerical model output, here we explicitly leverage EVT by specifying
that the conditional distribution of monitor data given the RFM has a
GPD tail. However, modeling only the extremes is not sufficient here,
since even the center of the conditional distribution of the monitor
data given an extremely large value of the RFM may in fact be extreme.
This requires that we construct a flexible model for the entire
conditional distribution which uses EVT to characterize the upper tail.
Our overall model is a combination of quantile regression and EVT,
employing quantile regression at levels where there is adequate
information to fit a flexible model and EVT in the tail, which allows
one to extrapolate beyond the range of the data. We assume that EVT is
appropriate for the upper tail of the conditional distribution given
any value of the RFM (high or low) and allow the GPD distribution to
vary with the RFM and spatial location. Our work is similar in spirit
with a recent study by \citet{Bentzien-2012}, who use a related
strategy for probabilistic quantitative precipitation forecasting. They
estimate a conditional mixture with a GPD tail, although without
spatial variation in the parameters or a RFM with unknown input
parameters. Third, rather than using the standard diagnostic tools to
select a threshold, our model estimates the threshold above which EVT
becomes appropriate. Prior work on threshold estimation is limited;
\citet{Frigessi-2003} use a mixture of a parametric light-tailed
distribution and a GPD, and \citet{behrens-2004a} use a
semi-parametric center and GPD tail and estimate the threshold in a
Bayesian way.

\section{Description of the air quality model and monitor data}\label{sdata}

\subsection{Base CMAQ model}
The Community Multiscale Air Quality (CMAQ) model [\citet{Byun2006}],
version 4.7.1 [\citet{Foley2010}], is chosen as the regional
photochemical transport model used as the base simulation, from which
we later construct our RFM. Ozone was simulated hourly with CMAQ in a
domain centered on the southeastern United States for an episode
between July 1, 2005 and September 30, 2005, with the full month of
June 2005 as a spin-up period. Eight-hour average ozone is then
computed using the hourly values. Standard model configuration was used
with a 12~km by 12~km horizontal grid spacing and 14 vertical layers from
the surface to 100 hPa, and the Statewide Air Pollution Research Center
(SAPRC99) gas-phase chemical mechanism [\citet{Carter2000}].
Meteorological fields were developed using the fifth generation
mesoscale model (MM5), version 3.6.3 [\citet{Grell1995}], and chemical
emissions based on the 2001 National Emissions Inventory
(\href{http://www.epa.gov/ttn/chief/emch/index.html\#2001}{http://www.epa.gov/ttn/chief/emch/}
\href{http://www.epa.gov/ttn/chief/emch/index.html\#2001}{index.html\#2001}) were processed
using the SMOKE processor, version 2.3.2 (\href{http://www.smoke-model.org}{http://www.smoke-model.org}),
augmented with year 2005 specific emissions data for electric generating units equipped with Continuous
Emission Monitoring Systems (CEMS), mobile emissions processed by
MOBILE~6 (\href{http://www.epa.gov/otaq/m6.htm}{http://www.epa.gov/otaq/m6.htm}), and meteorologically
adjusted biogenic emissions from the Biogenic Emission Inventory System
(BEIS) 3.13 [\citet{Schwede2005}].

\subsection{Reduced-form CMAQ model}\label{sRFM}

When model runs are computationally intensive and many runs are
required for a thorough sensitivity analysis, an approximation to the
model output can be used. One such technique is the decoupled direct
method in three dimensions (DDM-3D). This gives a reduced-form CMAQ
(RF-CMAQ) model described below. Eulerian photochemical models such as
CMAQ typically simulate the emissions, transport, and chemistry of
gases and particles in the atmosphere by numerically solving the
atmospheric diffusion equation [\citet{Seinfeld98}]
%
\begin{equation}
\label{ADE} \frac{\partial{C_i}}{\partial t} = -\btriangle\cdot (\bu C_i ) +
\btriangle\cdot (\bK\btriangle C_i ) + R_i +
E_i,
\end{equation}
where $C_i(t,\bs)$ is the concentration of species $i= 1,2,\ldots,N$ at
time $t$ and location~$\bs$ (with notation for space and time dropped
for simplicity), $\bu$ is fluid velocity, $\bK$ is the diffusivity
tensor, $R_i$ is the net rate of chemical generation of species, and
$E_i$ is the species emissions rate. DDM-3D computes first-order
semi-normalized sensitivity coefficients $S_{ij}^{(1)}(t,\bs)$ to
perturbations in an input parameter $p_j$ as
%
\begin{equation}
\label{S1} S_{ij}^{(1)} = \frac{\partial C_i}{\partial\varepsilon_j},
\end{equation}
where $\varepsilon_j$ is a scaling variable with a nominal value of
1.0 applied to the unperturbed parameter field, $\tilde{p}_j$ as
$\varepsilon_j = \frac{p_j}{\tilde{p}_j}$. Differentiating (\ref
{ADE}) while using the above definitions leads to an analogous equation
governing the first-order sensitivity field
%
\begin{equation}
\label{SEN} \frac{\partial{S_{ij}}}{\partial t} = -\btriangle\cdot (\bu S_{ij} ) +
\btriangle\cdot (\bK\btriangle S_{ij} ) + J_iS_{j}
+ {\tilde E}_{i},
\end{equation}
where ${\tilde E}_{i}$ is the unperturbed emission rate, and $\bJ_i$
is the $i$th row vector in the Jacobian matrix $\bJ$
($J_{ij}=\partial{R_i}/\partial{C_j}$) representing the chemical
interaction between species as the previously undefined terms.

\begin{figure}

\includegraphics{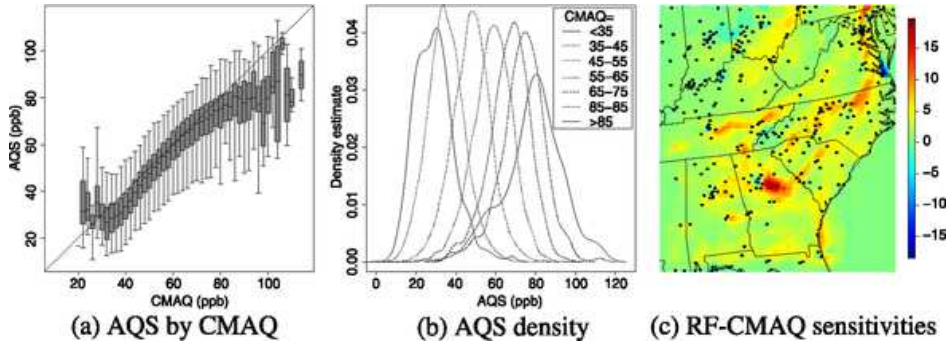}

\caption{Plot of the CMAQ output versus AQS monitor data pooled over
all sites [panel~\textup{(a)}], the kernel-smoothed density estimate
of the AQS
density by binned CMAQ [panel~\textup{(b)}], and the RF-CMAQ sensitivity
$S_1^{(1)}(1,\bs)$ for one day for mobile source NO$_x$ [panel~\textup{(c)};
points are AQS monitor locations].}\label{fdata}
\end{figure}

Calculations for second-order sensitivity coefficients are also
possible and are defined as
%
\begin{equation}
\label{S2} S_{ijk}^{(2)}=\frac{\partial^{2}{C_{i}}}{\partial\varepsilon_j\,
\partial\varepsilon_k}.
\end{equation}
For a given species, Taylor series expansion can be used to approximate
the concentration $C_i(t, \bs)$ as a function of perturbations in a
set of input parameters of interest using first- and second-order
sensitivity coefficients [\citet{Cohan2005}]. In this application we
are interested only in the concentration of a single species, ozone,
and, therefore, the subscript $i$ is dropped. The RF-CMAQ model for $d$
parameters including second-order and cross-sensitivities is
%
\begin{eqnarray}
\label{C} C(t,\bs|\balpha) &=& C_{0}(t,\bs) + \sum
_{j=1}^dS^{(1)}_{j}(t,\bs )
\alpha_j + \frac{1}{2}\sum_{j=1}^dS^{(2)}_{jj}(t,
\bs)\alpha_j^2
\nonumber
\\[-8pt]
\\[-8pt]
\nonumber
&&{} + 0.5\sum_{l\ne j}S^{(2)}_{lj}(t,
\bs)\alpha_j\alpha_l,
\end{eqnarray}
where $C(t,\bs|\balpha)$ is the ozone concentration due to a specific
set of perturbations $\balpha=(\alpha_1,\ldots,\alpha_d)$ at time $t$
and location $\bs$, $C_{0}(t,\bs)$ is unperturbed concentrations from
the base CMAQ simulation, and $\alpha_j$ is the perturbation in input
parameter $p_j$. For example $\alpha_j=-0.10$ corresponds to 10\%
decrease in NO$_{\mathrm{x}}$ emissions compared to the NO$_{\mathrm{x}}$
emissions used for $C_0$. The sensitivity coefficients produced by
DDM-3D vary in space and time, providing a computationally efficient
calculation of ozone under different perturbations in emissions inputs
through the RFM. For example, in urban centers NO$_{\mathrm{x}}$
emissions frequently act as a sink of ozone resulting in negative
sensitivity to sectors involving NO$_{\mathrm{x}}$ emissions [Figure
\ref{fdata}(c)]. In this analysis we consider sensitivity to $d= 6$
inputs: mobile-source NO$_{\mathrm{x}}$ emissions (e.g., traffic),
point-source NO$_{\mathrm{x}}$ emissions (e.g., power plants), other
NO$_{\mathrm{x}}$ emissions (e.g., construction equipment),
anthropogenic VOCs emissions (e.g., benzene emitted from fuel
combustion by motor vehicles), biogenic VOCs emissions (e.g., limonene
emitted from pine trees), and ozone boundary conditions (3-D hourly
pollutant concentrations specified at the grid cells surrounding the
model domain).

This second-order RFM is used as an emulator for the full CMAQ model.
That is, we use (\ref{C}) to approximate the spatiotemporal output
that would result from an evaluation of the full CMAQ model for a new
set of perturbations $\balpha$. This RFM has been shown to have
normalized mean error within 10\% of reevaluating the full model for
$\balpha$ perturbations up to $-$100\% [\citet{Cohan2005}]. An important
caveat is that \citet{Cohan2005} did not address differences between
the full CMAQ and RFM for extreme values, which could potentially have
greater impact on projections of extremes.

Using the RF-CMAQ model provides substantial computational improvement.
A single month-long simulation using the full CMAQ model takes
approximately 10 hours using 72 processors on an IBM system x
iDataPlex. The DDM-3D model on the same system runs in approximately 10
days but gives us the ability to estimate pollutant concentrations
under many different emissions levels. In the paper, the Bayesian
framework allows us to evaluate the reduced form model at thousands of
different emissions perturbations to create posterior distributions of
the ozone concentration across space. In contrast, running the full
model just 100 times in order to roughly approximate the uncertainty in
the emissions inputs would require more than a month of computational
time. Therefore, RF-CMAQ is the only viable way of exploring the
effects of control strategies while accounting for uncertainty in the
emissions inputs.

\subsection{AQS monitor data}

Ozone predictions are also paired in time and space with hourly average
ozone observations obtained from EPAs Air Quality System (AQS;
\url{http://www.epa.gov/ttn/airs/airsaqs/}). This analysis focuses on the
maximum eight-hour average ozone concentrations per day (MD8 O$_3$) for
July 1 to September 30 in 2005 at 307 monitoring stations in the
southeastern US in Figure~\ref{fdata}(c). The MD8 is the averaging
metric of interest, because it is used for determining compliance with
the EPA's National Ambient Air Quality Standards for ozone.

\section{Statistical model for extreme ozone}\label{smodel}

Let $y(t,\bs)$ be the AQS measurement for day $t\in\{1,\ldots,n_T\}$ at
spatial location $\bs$. Our objective is to estimate the conditional
distribution of $y(t,\bs)$ given RF-CMAQ output for the grid cell
containing location~$\bs$, denoted $C(t,\bs|\balpha)$. As described
in Section \ref{sRFM}, the perturbation vector $\balpha$ is treated
as an unknown parameter in the hierarchical model to allow the AQS data
to determine the optimal adjustment to the initial emission levels. For
notational convenience, we temporarily suppress dependence on $t$, $\bs
$ and $\balpha$ and simply describe the model for $y$ given $C$. We
specify a flexible semiparametric model below a threshold $\mu$ where
data are abundant, and transition to a parametric GPD model above a
threshold where data are sparse. The models above and below the
threshold are described in Sections \ref{sGP} and \ref{squantreg},
respectively.

\subsection{Parametric EVT model above the threshold}\label{sGP}
The parametric GPD distribution has three parameters: lower bound $\mu
$, scale $\sigma>0$, and shape $\xi$. The domain is $(\mu,\infty)$
if $\xi>0$ and $(\mu, \mu-\sigma/\xi)$ if $\xi<0$, and the
density and quantile (inverse CDF) functions are
%
\begin{eqnarray*}
\label{GP} \operatorname{dGPD}(y|\mu,\sigma,\xi) &=& \frac{1}{\sigma} \biggl(1+
\frac{\xi
}{\sigma}(y-\mu) \biggr)_+^{-1/\xi-1} \quad\mbox{and}\\ \operatorname{qGPD}(\tau|
\mu,\sigma,\xi)& =& \mu+\frac{\sigma}{\xi} \bigl([1-\tau]^{-\xi}-1 \bigr),
\end{eqnarray*}
respectively, where $x_+ = \max\{0,x\}$. In practice, typically a
threshold is selected to be the GPD lower bound, $\mu$. Unlike typical
extreme value analysis, rather than choosing a threshold, we instead
treat it as a parameter in the model-fitting process.

As shown in Figure \ref{fdata}, the distribution of AQS values is
highly dependent on the RF-CMAQ output. Therefore, we assume that the
semiparametric/parametric threshold depends on $C$. Since the threshold
will likely vary more on the data scale than the percentile scale, we
model the threshold and conditional distribution via its quantile
function. The conditional quantile function $q(\tau|C)$ satisfies
$P[y<q(\tau|C)] = \tau\in[0,1]$ and, therefore, the conditional
density function is $d q^{-1}(y|C)/dy$. Utilizing the GPD, the full
conditional quantile function is
%
\begin{equation}
\label{condquant1} q(\tau|C) = \cases{ %
\displaystyle q_0(\tau|C), &\quad $\tau\le T(C),$
\vspace*{2pt}\cr
\displaystyle \operatorname{qGPD} \biggl[\frac{\tau-T(C)}{1-T(C)}\Big|\mu(C),\sigma(C),\xi(C) \biggr], &\quad $\tau>
T(C).$}
\end{equation}
In this model, $T(C) \in[0,1]$ is the quantile level that separates
the semiparametric quantile function $q_0$ and the parametric $\operatorname{qGPD}$.
For $\tau$ above $T(C)$, and thus $y$ above $q_0[T(C)|C] = \mu(C)$,
the quantile function takes the form of a GPD with lower bound $\mu
(C)$, scale $\sigma(C)$, and shape $\xi(C)$.

\subsection{Semiparametric quantile regression below the
threshold}\label{squantreg}

We use the model of \citet{Reich-2012} for the quantile function below
the threshold, $q_0$. We assume that
%
\begin{equation}
\label{q0} q_0(\tau|C) = \beta(C) + \sum
_{l=1}^L B_l(\tau)\theta_l(C).
\end{equation}
The quantile function is the sum of an overall location term $\beta
(C)$ and a linear combination of known basis functions $B_l$ with
unknown coefficients $\theta_l(C)$ which determine the shape of the
quantile function given $C$. For the choice of basis functions below,
$\beta(C)$ is the median. Although this model is quite flexible, it is
centered on the heteroskedastic Gaussian model in that if $\theta
_1(C)=\cdots=\theta_L(C)$, then the quantile function reduces to the
Gaussian quantile function with mean $\beta(C)$ and standard deviation
$\theta_1(C)$.

To lead to a valid statistical model, $q_0$ must be increasing in $\tau
$ for all $C$. To do this, we define $B_1(\tau) = \Phi^{-1}(\tau)$
if $L=1$, where $\Phi^{-1}$ is the standard normal quantile function.
In this case, the model below the threshold is Gaussian with mean
$\beta(C)$ and standard deviation $\theta(C)$. For non-Gaussian data
we generalize by allowing $L\ge2$ and specifying basis functions
%
\begin{equation}
\label{B1} B_l(\tau) = \cases{ %
\Phi^{-1}(\kappa_{l}) - \Phi^{-1}(
\kappa_{l+1}), & \quad $\tau<\kappa_{l},$
\vspace*{2pt}\cr
\Phi^{-1}(\tau) - \Phi^{-1}(\kappa_{l+1}), &\quad
$\kappa_{l}\le\tau <\kappa_{l+1},$
\vspace*{2pt}\cr
0, & \quad $\kappa_{l+1}\le\tau,$}
\end{equation}
for $l$ with $\kappa_l<0.5$ and
%
\begin{equation}
\label{B2} B_l(\tau) = \cases{ %
0, &\quad
$\tau<\kappa_{l},$
\vspace*{2pt}\cr
\Phi^{-1}(\tau) - \Phi^{-1}(\kappa_{l}), &
\quad $\kappa_{l}\le\tau <\kappa_{l+1},$
\vspace*{2pt}\cr
\Phi^{-1}(\kappa_{l+1}) - \Phi^{-1}(
\kappa_{l}), &\quad $\kappa_{l+1}\le \tau,$}
\end{equation}
for $l$ such that $\kappa_l\ge0.5$, where $0=\kappa_1<\cdots<\kappa
_{L+1}=1$ is a grid of equally-spaced knots covering $[0,1]$. Then the
quantile function is increasing if and only if $\theta_{l}(C)>0$ for
all $l$ and $C$. We only consider even $L$ in which case $B_l(0.5)=0$
for all $l$, and the median is $q_0(0.5|C)=\beta(C)$. Also, if $\theta
_{1}(C)=\cdots=\theta_{L}(C) = \theta(C)$, then for $\tau<\mu(C)$,
$q(\tau|C) = \beta(C) + \theta(C)\Phi^{-1}(\tau)$, and the density
below the threshold is Gaussian with mean $\beta(C)$ and standard
deviation $\theta(C)$. It is also possible to use the other basis
function $B_l$. For example, taking $B_l$ to be the gamma or log-normal
distribution function would ensure a lower bound $q_0(0|C)=0$.

These basis functions also permit a closed-form expression for the
conditional density
%
\begin{eqnarray}
\label{density}&& p(y|C)\nonumber \\
&&\qquad=I\bigl[y<\mu(C)\bigr]\sum
_{l=1}^L I\bigl[q_0(
\kappa_{l}|C)\le y < q_0(\kappa_{l+1}|C)\bigr]
\mathrm{N} \bigl[y|a_l(C),\theta_l(C)^2
\bigr]
\\
&&\qquad\quad{}+ I\bigl[y\ge\mu(C)\bigr] \bigl[1-T(C)\bigr]\operatorname{dGPD} \bigl(y|\mu(C),\sigma(C),\xi
(C) \bigr),\nonumber
\end{eqnarray}
where $\mathrm{N}(\cdot|a,b^2$) denotes the density of a normal with mean $a$
and standard deviation $b$, and $a_l(C) = q(\kappa_{l+1}|C) - \theta
_l(C)\Phi^{-1}(\kappa_{l+1})$ if $\kappa_l< 0.5$ and $a_l(C) =
q(\kappa_l|C) - \theta_l(C)\Phi^{-1}(\kappa_l)$ if $\kappa_l\ge
0.5$. Therefore, the density is multiply-split normal with breakpoints
(and points of discontinuity) $q(\kappa_l|C)$ and $\mu(C)$. However,
our primary interest, the quantile function, is a continuous function.

\subsection{Modeling dependence on RF-CMAQ}

The conditional density varies with~$C$ via $\beta(C)$, $\theta
_l(C)$, $T(C)$, $\sigma(C)$, and $\xi(C)$. These parameters could be
allowed to have a complex dependence on $C$ to capture subtle features
of the conditional distribution. For example, one could use a Gaussian
process defined over~$C$. For simplicity, we assume that after a
suitable transformation each parameter is an order-$M$ polynomial
expansion of $C$. That is, $\beta(C) = \bX\ba^{(\beta)}$,
$\log[\theta_l(C)] = \bX\ba^{(\theta_l)}$, $\log[\sigma(C)] =
\bX\ba^{(\sigma)}$, $\xi(C) = \bX\ba^{(\xi)}$, where $\bX=
(1,{\bar C},\ldots,{\bar C}^M)$ and ${\bar C}=(C-50)/15$ is the
standardized CMAQ output (where 50 and 15 are the approximate mean and
standard deviation, resp.). Note that this polynomial model
contains the linear regression model as a special case and that in this
case the intercept can account for systematic bias between the AQS data
and the RF-CMAQ predictions. Higher-order polynomials or spline basis
expansion would allow for more complex relationships between RF-CMAQ
and the AQS data.

The semi-parametric/parametric threshold $T(C)$ must be modeled so that
$T(C)$ is confined to $[0,1]$ for all $C$. In our analysis, we intend
for $T(C)$ to be an extreme quantile to theoretically justify the GPD
fit, so we restrict $T(C)\in[l,u]$ where $l$ and $u$ are unknown
parameters with $l\sim \operatorname{Uniform}(0.8,1.0)$ and $u|l\sim
\operatorname{Uniform}(l,1.0)$. The variability of $T(C)$ within $(l,u)$ is modeled
using the logistic link
%
\begin{equation}
\label{c} T(C) = l\frac{\exp[d(C)]}{1+\exp[d(C)]} +u\frac{1}{1+\exp[d(C)]}.
\end{equation}
As with the other parameters, $d$, and thus $T$, varies with $C$ as
$d(C) = \bX\ba^{(d)}$.


\subsection{Spatiotemporal modeling}\label{smodel3}

There are two potential sources of spatial and temporal dependence in
the data: spatial variation in the conditional distribution of AQS
given RF-CMAQ, and residual spatiotemporal association in the
observations given the conditional distribution. To account for spatial
variation in the conditional distribution, we allow the parameters that
define the semiparametric model below the threshold, $\ba^{(\beta)}$
and $\ba^{(\theta_l)}$, as well as the GPD scale $\ba^{(\sigma)}$
to vary by spatial location. These processes are then smoothed by
Gaussian process priors. For example, denote $\ba^{(\beta)}$ at
location $\bs$ as $\ba^{(\beta)}(\bs) = [a_0^{(\beta)}(\bs
),\ldots,a_M^{(\beta)}(\bs)]^T$ and, thus, the spatially-varying
coefficients $\beta(x,\bs) = \bX\ba^{(\beta)}(\bs)$. Then
$a_j^{(\beta)}(\bs)$ has a Gaussian process prior with mean ${\bar
a}_j^{(\beta)}$, variance $\tau_j^{(\beta)}$, and exponential
spatial correlation $\operatorname{Cor}[a_j^{(\beta)}(\bs),a_j^{(\beta)}(\bs')] =
\exp(-\Vert \bs-\bs'\Vert /\rho)$. The hyperparameters have priors ${\bar
a}_j^{(k)}\sim \mathrm{N}(0,c_1^2)$ and $\tau_j^{(k)}\sim \operatorname{Gamma}(c_2,c_3)$
for $k\in\{\beta,\theta_1,\ldots,\theta_L,\sigma\}$. To borrow
strength across processes, we assume a common spatial range $\rho$,
which is reasonable since all of these spatially-varying parameters
represent the change in response distribution across locations.

We find that while the threshold parameters $\ba^{(d)}$ and GPD shape
parameters $\ba^{(\xi)}$ are well identified when allowed to vary by
the value of RF-CMAQ, they are poorly identified when allowed to vary
by RF-CMAQ and spatial location. This is not surprising since they are
by definition related only to the tail of the distribution and, thus,
there are only a few relevant observations at each location. These
parameters are thus held constant for all locations with priors
$a_j^{(k)}\sim \mathrm{N}(0,c_1^2)$ for $k\in\{d,\xi\}$. We note that
although the threshold is fixed at a constant quantile level across
space, the actual threshold on the ozone scale is $\mu[C(t,\bs
|\balpha)]=q_0\{T[C(t,\bs|\balpha)]|C(t,\bs|\balpha),\bs\}$,
which does vary spatially. Also, since the interpretation of the GPD
scale is dependent on the threshold, it seems a reasonable approach to
not allow the parameters dictating the threshold to vary spatially and
to assume that the spatially varying scale parameter can account for
spatial variation. Combining these specifications gives the final
quantile model fit to the ozone data in Section \ref{sanalysis},
%
\begin{equation}
\label{condquantspace}\quad q(\tau|C,\bs) = \cases{ %
\displaystyle q_0\bigl[\tau|C(t,\bs|\balpha),\bs\bigr], \qquad \tau\le T\bigl[C(t,\bs|
\balpha)\bigr],
\vspace*{2pt}\cr
\displaystyle \operatorname{qGPD} \biggl\{\frac{\tau-T[C(t,\bs|\balpha)]}{1-T[C(t,\bs|\balpha)]}\Big| \mu\bigl[C(t,\bs|\balpha)\bigr], \sigma
\bigl[C(t,\bs|\balpha),\bs\bigr],\vspace*{2pt}\cr
\hspace*{185pt}\xi\bigl[C(t,\bs|\balpha)\bigr] \biggr\},\vspace*{2pt}\cr
\hspace*{105pt}\tau>
T\bigl[C(t,\bs|\balpha)\bigr],}
\end{equation}
where $q_0[\tau|C(t,\bs|\balpha),\bs] = \beta[C(t,\bs|\balpha
),\bs] + \sum_{l=1}^L B_l(\tau)\theta_l[C(t,\bs|\balpha),\bs]$.

Even after accounting for spatial variation in the conditional
distributions, there is spatial and temporal dependence in the
residuals due to day-to-day variation in ozone. We account for this
dependence with a Gaussian copula [\citet{Nelson99}]. The copula is
defined by a latent Gaussian process $z(t,\bs)$ with mean zero,
variance one, and a spatiotemporal correlation function. Then $U(t,\bs
)=\Phi[z(t,\bs)]\sim \operatorname{Unif}(0,1)$, and the latent process is related
to the response as $y(t,\bs) = q[U(t,\bs)|C(t,\bs|\balpha),\bs]$.
While the Gaussian copula induces some spatiotemporal dependence, it is
well known that the Gaussian copula gives asymptotic independence. That
is, assuming the same marginal distribution for $y(t,s)$ and $y(t',\bs
')$, then $\lim_{u\rightarrow U}P[y(t,s)>u|y(t',\bs')>u] = 0$, where
$U$ is the upper bound of $y(t,\bs)$, which implies that the Gaussian
copula is equivalent to ignoring dependence between very extreme
events. Therefore, the Gaussian copula may not be ideal for extreme
data in all settings. However, our exploratory analysis in Section~\ref
{sanalysis} suggests that there is little extremal dependence in the
residuals and, therefore, that this model fits the AQS data well after
accounting for RF-CMAQ output. In other cases, copulas with asymptotic
dependence may be desirable. Examples of copulas with asymptotic
dependence include the t-copula [\citet{Nelson99}] or a nonparametric
copula [\citet{fuentes-2010a}]. Another possibility is to specifically
target extremal dependence [e.g., \citet
{davison-1990,chavez-2012,eastoe-2012}].

\section{Computational approach to evaluating control
strategies}\label{scomp}

The computing used for the ozone data analysis has two main steps: we
first analyze the AQS data to estimate the parameters in the
conditional distribution of AQS given RF-CMAQ, and then generate
replications of the summer ozone process under different control
strategies. These two steps are described separately in the subsections below.

\subsection{Parameter estimation}\label{sMCMC}

Because of the size of the data set, we fit this model in two stages.
We first estimate $\balpha$ and the $\ba^{(k)}$ parameters for $k\in
\{\beta,\theta_l,\sigma,\xi,d\}$ in one model fit assuming the
observations are independent conditioned on these parameters. In the
second stage, we estimate the copula parameters conditioned on the
first-stage parameter estimates. Assuming independence of the
observations conditioned on $\Theta(\bs) = \{\balpha, \ba^{(\beta
)}(\bs),\ba^{(\theta_l)}(\bs),\ba^{(\sigma)}(\bs),\ba^{(\xi
)},\ba^{(d)},l,u\}$, the likelihood is simply the product of terms of
the form of (\ref{density}). Denote $p[y(\bs,t)|\Theta(\bs)]$ as
the density of $y(\bs,t)$. None of the parameters in the likelihood
have conjugate full conditionals, so we use Metropolis--Hastings
sampling. Metropolis--Hastings sampling proceeds by specifying initial
values for all parameters and then updating the parameters
one-at-a-time, conditioned on the current value of all other
parameters. For example, to update $\alpha_j$, we draw candidate
$\alpha_j^{\mathrm{can}} \sim \mathrm{N}(\alpha_j^{\mathrm{cur}},c^2)$, where $\alpha
_j^{\mathrm{cur}}$ is the current value and the standard deviation $c$ is a
tuning parameter. With probability $R$, $\alpha_j$ is set to $\alpha
_j^{\mathrm{can}}$, and $\alpha_j$ is set to $\alpha_j^{\mathrm{cur}}$ otherwise, where
\[
R = \min \biggl\{ \frac{\prod_{i,t}p[y(\bs_i,t)|\Theta(\bs_i)^{\mathrm{can}}]p(\alpha_j^{\mathrm{can}})} {
\prod_{i,t}p[y(\bs_i,t)|\Theta(\bs_i)^{\mathrm{cur}}]p(\alpha
_j^{\mathrm{cur}})},1 \biggr\},
\]
$\Theta(\bs)^{\mathrm{can}}$ includes $\alpha_j^{\mathrm{can}}$, $\Theta(\bs)^{\mathrm{cur}}$
includes $\alpha_j^{\mathrm{cur}}$, and $p(\alpha)$ is the Gaussian prior.
Evaluating $p[y(\bs,t)|\Theta(\bs)]$ requires first computing
RF-CMAQ given perturbation parameters $\balpha$, $C(\bs,t|\balpha)$,
which is trivial for the RF-CMAQ model following (\ref{C}). All
parameters are updated similarly with Gaussian candidate distributions
tuned to give acceptance rates near 0.4. We generate 25,000 samples
from the posterior and discard the first 10,000 as burn-in. Convergence
is monitored using trace plots of several representative parameters.

To estimate the copula parameters, we compute the estimated
Gaussian-transformed residuals $z(t,\bs) = \Phi^{-1} \{{\hat
q}^{-1}[y(t,s)|C(t,\bs),\bs] \}$, where ${\hat q}$ is the
quantile function evaluated at the posterior mean of all model
parameters. There is of course spatial dependence in the residuals, and
sampling with spatial dependence would be crucial for statistics
defined over the spatial domain, for example, total precipitation in a
watershed. However, our interest is in projecting the change in ozone
distribution at each site, and not for a collection of sites
simultaneously. Therefore, assessment of spatial dependence in the
predictions is not a concern and we assume the residuals are
independent over space for computational convenience. We then fit a
first-order autoregressive model for the temporal dependence,
$\operatorname{Cor}[z(t,\bs),z(t',\bs')] = \exp(-|t-t'|/\phi)I(\bs=\bs')$, with
autocorrelation parameters held constant over space and time. We fix
$\phi$ to match the sample correlation of subsequent residuals at the
same location.

\subsection{Generating samples from the posterior predictive
distribution} \label{spredict}

To evaluate the effects of control strategies on the likelihood and
magnitude of extreme ozone events, we generate several replications of
summer ozone at each spatial location. The control strategies
correspond to reductions of emissions in various sectors and are
parameterized in terms of the RF-CMAQ model inputs $\balpha$. In the
RF-CMAQ model, $\alpha_j$ represents a $100\alpha_j$\% change in the
initial estimate of the emissions in sector $j$. Therefore, the
posterior of $\balpha$ represents the calibrated emissions for the
base case based on fitting to the AQS data. To simulate RFM output that
corresponds to an additional change of $100\eta_j$\% after the
calibration via $\alpha_j$, we use $\alpha^*_j = (1+\alpha
_j)*(1+\eta_j) - 1$ as inputs to the RFM. By simulating data for
different values of $\boldeta=(\eta_1,\ldots,\eta_p)$, we simulate
ozone data under different control strategies. These control strategies
assume a uniform reduction across the entire region.

For each control strategy we generate $R=10\mbox{,}000$ replicates of the
summer ozone at each CMAQ grid cell. Due to the computational burden,
the grid cells are thinned by removing every other column and row. For
each replicate we randomly sample one of the posterior draws for the
parameters in the conditional distribution $\{\ba^{(\beta)}, \ba
^{(\theta_l)}, \ba^{(d)}, \ba^{(\sigma)}, \ba^{(\xi)}, \balpha\}$. At each iteration all spatially-varying parameters are interpolated
from the AQS stations used for model-fitting to the grid cell locations
by sampling from their posterior predictive distribution. We then
compute the RF-CMAQ model corresponding to input $\balpha^*=(\alpha
^*_1,\ldots,\alpha^*_p)$, and the conditional distribution of $y(t,\bs)$
given $C(t,\bs|\balpha^*)$ and $\{\ba^{(\beta)}, \ba^{(\theta
_l)}, \ba^{(d)}, \ba^{(\sigma)}, \ba^{(\xi)}\}$. We generate the
responses for each simulated year for location $\bs$ by generating
$\bz(\bs) = [z(1,\bs),\ldots,z(n_T,\bs)]^T$ from a multivariate normal
model with mean zero and covariance $\operatorname{Cov}[z(t,\bs),z(t',\bs)] = \exp
(-\Vert t-t'\Vert /\phi)$ and transforming to $y(t,\bs) = q(\Phi\{z(t,\bs)\}
|C(t,\bs|\balpha^*,\bs))$ so that $y(t,\bs)$ has the quantile
function in~(\ref{condquantspace}).

This method of simulation accounts for uncertainty in the AQS value
given RFM output, and uncertainty in the parameters in the conditional
distribution of the AQS value given the RFM output. It also partially
accounts for randomness in the RF-CMAQ output by marginalizing over the
posterior of $\balpha$. However, there are many additional inputs to
RF-CMAQ that are taken as fixed, and so not all the randomness in
RF-CMAQ from year to year is accounted for by this approach. Ideally,
we would have a larger sample of RF-CMAQ output to better represent the
sampling distribution of RF-CMAQ from year to year. Therefore, the
results should be interpreted cautiously as pertaining to the changes
in the ozone distribution for this particular simulated year, which may
suppress some variability for an arbitrary future year.

\section{Constructing the downscaler between RFM output and AQS
data}\label{sanalysis}

To display the results of fitting the conditional distribution of AQS
given RF-CMAQ, we first compare several models based on test set
prediction in Section \ref{scompare}. We then illustrate the fitted
distribution of our final model in Section \ref{sfitted}.

\subsection{Model comparisons}\label{scompare}

We compare several models by varying the number of basis functions in
the semiparametric quantile process, $L$, the order of the polynomial
for RF-CMAQ in the conditional distribution, $M$, and with and without
[i.e., $T(C)=1$] the GPD tail. For comparison, we also include the
nonstatistical forecast by simply taking the base CMAQ output $C_0(\bs,t)$ as the prediction. For all fits, we use uninformative priors
$c_1=100$, $c_2=c_3=0.1$, and $\log(\rho) \sim \mathrm{N}(0,10)$. To
compare these models, we randomly (across space and time) split the
data equally into training ($n=13\mbox{,}645$) and testing data sets
($n=13\mbox{,}645$). We fit each model to the training set, calculate the
posterior mean of all model parameters, and then compute the predictive
distribution for each test set observation.

Models are compared in terms of their fit to the upper tail of the
distribution using Brier scores for exceedances and quantile scores for
extreme quantiles [see, e.g., \citet{Gneiting-2007}]. The quantile
score for quantile level $\tau$ is $2 \{I[y<{\hat q}(\tau)]-\tau
\}({\hat q}-y)$, where $y$ is the test set AQS value and ${\hat
q}(\tau)$ is its estimated $\tau${th} quantile. The Brier score for
evaluating accuracy of predicting exceedance of threshold $c$ is
$[e(c)-P(c)]^2$, where $e(c) = I(y>c)$ is the indicator that the test
set AQS value exceeds $c$ and $P(c)$ is the predicted probability of an
exceedance. For the nonstatistical predictions, we take ${\hat q}(\tau
) = C_0$ and $P(c) = I(C_0>c)$. We compare models using several extreme
values of $\tau$ and $c$, and average these values over all
observations in the test set. For both quantile and Brier scores, small
values are preferred.

\begin{table}
\def\arraystretch{0.9}
\tabcolsep=0pt
\caption{Quantile and Brier scores for various models.
Models vary by the number of basis functions in the quantile process
($L$), the degree of polynomial expansion of the RF-CMAQ predictors
($M$), and whether the upper tail is (GPD) or is not (NoGPD) a
generalized Pareto distribution. ``SLR,'' ``QR,'' and ``LR'' are simple
linear regression, quantile regression, and logistic regression,
respectively, with linear predictor $a(\bs)+b(\bs)C_0(\bs,t)$, where
$a(\bs)$ and $b(\bs)$ are estimated separately by site and $C_0(\bs,t)$ is the full CMAQ output. The lowest value (including ties) for
each criteria are in bold}\label{tCV}
\begin{tabular*}{\textwidth}{@{\extracolsep{\fill
}}ld{1.2}d{1.3}d{1.3}d{1.3}d{1.3}d{1.3}d{1.3}d{1.3}d{1.3}d{1.3}@{}}
\multicolumn{11}{c}{\textbf{(a) Quantile scores (ppb)}}\\
\hline
&&& \multicolumn{4}{c}{$\bolds{L=1}$} & \multicolumn
{4}{c@{}}{$\bolds{L=4}$}
\\[-6pt]
&&& \multicolumn{4}{c}{\hrulefill} & \multicolumn
{4}{c@{}}{\hrulefill} \\
 & & & \multicolumn
{2}{c}{$\bolds{M=1}$} & \multicolumn{2}{c}{$\bolds{M=2}$} &
\multicolumn{2}{c}{$\bolds{M=1}$} & \multicolumn{2}{c@{}}{$\bolds{M=2}$}
\\[-6pt]
& & & \multicolumn{2}{c}{\hrulefill} & \multicolumn{2}{c}{\hrulefill
} &
\multicolumn{2}{c}{\hrulefill} & \multicolumn{2}{c@{}}{\hrulefill}
\\
\multicolumn{1}{@{}l}{\multirow{2}{35pt}[10pt]{\textbf{Quantile level}}} & \multicolumn{1}{c}{\textbf{SLR}} & \multicolumn
{1}{c}{\textbf{QR}} & \multicolumn{1}{c}{\textbf{NoGPD}} &
\multicolumn{1}{c}{\textbf{GPD}} & \multicolumn{1}{c}{\textbf
{NoGPD}} & \multicolumn{1}{c}{\textbf{GPD}} & \multicolumn
{1}{c}{\textbf{NoGPD}} &
\multicolumn{1}{c}{\textbf{GPD}} & \multicolumn{1}{c}{\textbf
{NoGPD}} &
\multicolumn{1}{c@{}}{\textbf{GPD}} \\
\hline
0.750 & 7.73 & 5.51 & 5.37 & 5.35 & 5.33 & \multicolumn{1}{c}{\textbf
{5.30}\phantom{0}} & 5.35 & 5.35 &
5.31 & 5.32\\
0.950 & 7.66 & 2.02 & 1.85 & 1.84 & 1.84 & \multicolumn{1}{c}{\textbf
{1.81}\phantom{0}} & 1.84 & 1.84 &
1.83 & 1.83\\
0.990 & 7.66 & 0.768 & 0.521 & 0.507 & 0.518 & \multicolumn
{1}{c}{\textbf{0.498}} & 0.520 &
0.505 & 0.515 & 0.502\\
0.995 & 7.64 & 0.600 & 0.302 & 0.287 & 0.299 & \multicolumn
{1}{c}{\textbf{0.284}} & 0.301 &
0.287 & 0.299 & 0.289\\
\hline
\end{tabular*}
\vspace*{9pt}
\begin{tabular*}{\textwidth}{@{\extracolsep{\fill
}}ld{1.3}d{1.3}d{1.3}d{1.3}d{1.3}d{1.3}d{1.3}d{1.3}d{1.3}d{1.3}@{}}
\multicolumn{11}{c}{\textbf{(b) Brier scores (multiplied by 100)}}\\
\hline
&&& \multicolumn{4}{c}{$\bolds{L=1}$} & \multicolumn{4}{c}{$\bolds
{L=4}$} \\[-6pt]
&&& \multicolumn{4}{c}{\hrulefill} & \multicolumn
{4}{c@{}}{\hrulefill} \\
&& & \multicolumn{2}{c}{$\bolds{M=1}$} & \multicolumn{2}{c}{$\bolds
{M=2}$} &
\multicolumn{2}{c}{$\bolds{M=1}$} & \multicolumn{2}{c}{$\bolds
{M=2}$} \\[-6pt]
& & & \multicolumn{2}{c}{\hrulefill} & \multicolumn{2}{c}{\hrulefill
} &
\multicolumn{2}{c}{\hrulefill} & \multicolumn{2}{c@{}}{\hrulefill}
\\
\textbf{Threshold} & \multicolumn{1}{c}{\textbf{SLR}} & \multicolumn
{1}{c}{\textbf{LR}} & \multicolumn{1}{c}{\textbf{NoGPD}} &
\multicolumn{1}{c}{\textbf{GPD}} & \multicolumn{1}{c}{\textbf
{NoGPD}} & \multicolumn{1}{c}{\textbf{GPD}} & \multicolumn
{1}{c}{\textbf{NoGPD}} &
\multicolumn{1}{c}{\textbf{GPD}} & \multicolumn{1}{c}{\textbf{NoGPD}}
& \multicolumn{1}{c@{}}{\textbf{GPD}} \\
\hline
\phantom{0}70 & 9.38 & 6.88 & 6.18 & 6.17 & 6.13 & \multicolumn
{1}{c}{\textbf{6.11}\phantom{0}} & 6.19 & 6.21 & 6.12
& 6.14\\
\phantom{0}75 & 5.91 & 4.35 & 3.89 & 3.89 & 3.79 & \multicolumn
{1}{c}{\textbf{3.76}\phantom{0}} & 3.90 & 3.94 & 3.80
& 3.81\\
\phantom{0}80 & 3.15 & 2.43 & 2.11 & 2.11 & 1.98 & \multicolumn
{1}{c}{\textbf{1.95}\phantom{0}} & 2.12 & 2.15 & 1.98
& 1.97\\
\phantom{0}85 & 1.554 & 1.061 & 0.985 & 0.999 & 0.866 & \multicolumn
{1}{c}{\textbf{0.852}} & 0.997 &
1.022 & 0.859 & \multicolumn{1}{c}{\textbf{0.852}}\\
\phantom{0}90 & 0.791 & 0.458 & 0.427 & 0.440 & 0.350 & 0.344 & 0.430
& 0.445 &
0.343 & \multicolumn{1}{c}{\textbf{0.341}}\\
\phantom{0}95 & 0.418 & 0.277 & 0.225 & 0.229 & 0.184 & \multicolumn
{1}{c}{\textbf{0.182}} & 0.226 &
0.229 & \multicolumn{1}{c}{\textbf{0.182}} & \multicolumn
{1}{c}{\textbf{0.182}}\\
100 & 0.198 & 0.127 & 0.098 & 0.098 & 0.078 & \multicolumn
{1}{c}{\textbf{0.077}} & 0.098 &
0.098 & 0.079 & 0.078\\
\hline
\end{tabular*}
\end{table}

Table \ref{tCV} gives the results. The nonstochastic bias-adjusted
base CMAQ fit (``SLR''), that is, $a(\bs)+b(\bs)C_0(\bs,t)$ where
$a(\bs)$ and $b(\bs)$ are fit using separate linear regressions at
each location, has the highest scores, verifying the need for
statistical calibration. We also fit simple linear quantile regression
(``QR,'' using the \texttt{quantreg} package in \texttt{R}) and logistic
regression (``LR'') with base CMAQ as a linear predictor separate by
site (for very extreme quantiles and threshold these methods had some
convergence problems, and we simply carried forward the estimates from
the next lowest quantile or threshold). Although these models do not
provide a means to generate ozone under different scenarios, they do
provide improved fit compared to linear regression.

The quantile scores clearly show the value of the GPD tail model.
Although the score values are hard to interpret, the scores for the
0.99 and 0.995 quantiles are universally lower than those for the model
without the GPD tail; the scores for the statistical models for the
0.99 quantile are 0.521, 0.518, 0.520, and 0.515 for the models without
GPD tail, compared to 0.507, 0.498, 0.505, and 0.502 for the model with
GPD tail. Exceedence prediction for high thresholds is not only
affected by the tail of the distribution, but also the center. For
example, Figure \ref{fdata} shows that most exceedences of 80~ppb
occur when CMAQ is large and that 80~ppb is in the center of the
conditional distribution for large CMAQ. Therefore, the GPD tail is not
the most influential factor for Brier score, but rather the most
important factor is accurate modeling of the relationship between AQS
and RF-CMAQ via the degree of the polynomial in the model parameters,
$M$. The Brier score for 80~ppb is 2.105, 2.108, 2.122, and 2.150 for
the linear models with $M=1$, compared to 1.975, 1.951, 1.978, and
1.973 for the quadratic models with $M=2$. More complex models for the
RF-CMAQ predictors such as higher-order polynomials (i.e., $M>2$) or
spline fits are also possible. We fit the model with GPD tails and
$L=1$ with $M=3$ and found a slight improvement for moderate quantile
levels but poor performance for the extremes, likely due to
overfitting. Therefore, we conclude a second-order polynomial is
sufficient for these data. Non-Gaussian modeling ($L=4$) of the
distribution below the threshold does not appear to improve model fit
compared to the Gaussian model ($L=1$) for these data. Therefore,
although other models are fairly similar, the best model in terms of
both the quantile scores and Brier scores has $L=1$, $M=2$, and GPD
tail. This model is Gaussian below the threshold, and all the model
parameters are quadratic in RFM. The results below are from the data
analysis using this model on the complete data set.

\subsection{Summary of the final model}\label{sfitted}

For the final model with $L=1$ and $M=2$, RF-CMAQ input parameters
$\alpha_j$ are all negative with high probability, suggesting that all
emissions inputs used in the base simulation $C_0$ are too high. Their
95\% posterior intervals are ($-0.17, -0.05$) for mobile source
NO$_{\mathrm{x}}$, ($-0.22, -0.11$) for point source NO$_{\mathrm{x}}$,
($-0.27, -0.10$) for other NO$_{\mathrm{x}}$, ($-0.74, -0.59$) for
anthropogenic VOC emissions, ($-0.24, -0.17$) for biogenic VOC emissions,
and ($-0.10, -0.07$) for ozone boundary conditions.

\begin{figure}

\includegraphics{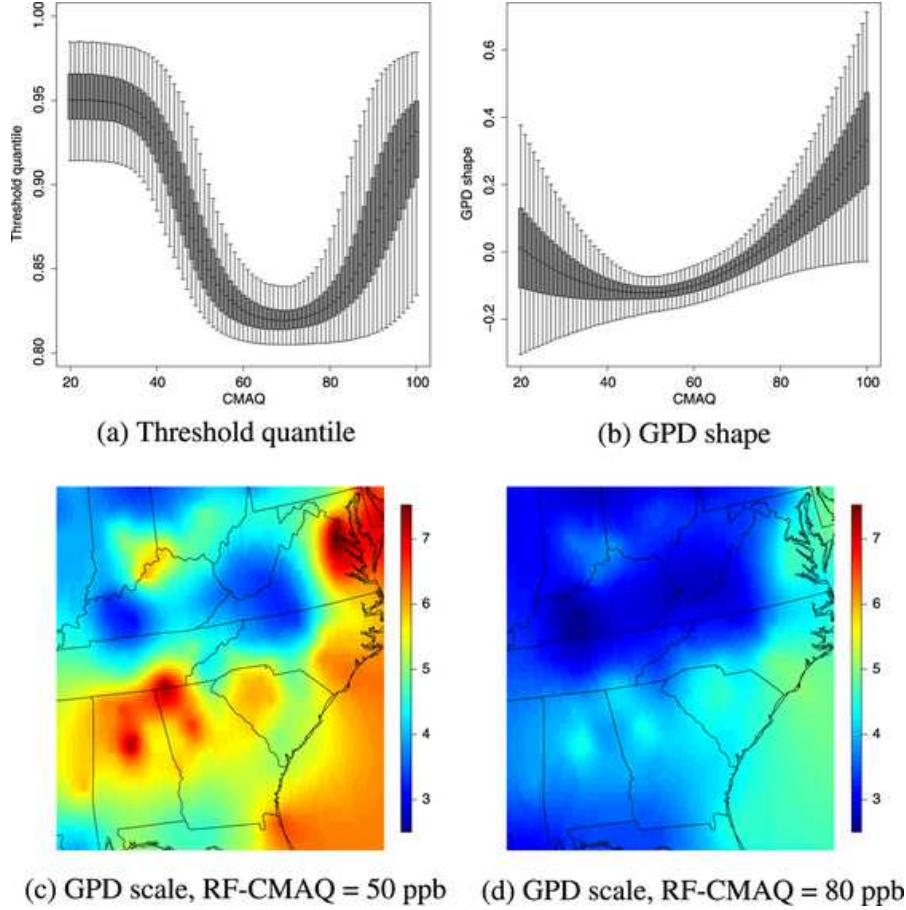}

\caption{Posterior of the threshold $T(C)$, GPD shape $\xi(C)$, and
GPD scale $\sigma(C,\bs)$ by RF-CMAQ output. In panels \textup{(a)}
and \textup{(b)},
the horizontal line in each boxplot gives the median, interquartile
range, and 95\% interval. Panels \textup{(c)} and \textup{(d)} plot
the posterior
mean.}\label{fGPD}
\end{figure}

Figure \ref{fGPD} summarizes the GPD fit to the tail of the
conditional distribution. The threshold $T[C(t,\bs|\balpha)]$ in
Figure \ref{fGPD}(a), which depends on $l$, $u$, and $d(C)$, varies
between the 0.80 and 0.95. The threshold is lower and, thus, the GPD
fits a larger portion of the tail, for moderate to high RF-CMAQ values
50--80~ppb. The GPD shape $\xi[C(t,\bs|\balpha)]$ in Figure \ref
{fGPD}(b) is near zero for low RF-CMAQ values, negative for moderate
RF-CMAQ, and positive for large RF-CMAQ. This generally agrees with the
sample density estimates in Figure \ref{fdata}(b), which have heavier
tails for low and high values of RF-CMAQ. Unlike the threshold and
shape, the GPD scale $\sigma[C(t,\bs|\balpha),\bs]$ varies
spatially [Figures \ref{fGPD}(c) and \ref{fGPD}(d)]. The GPD scale is
larger in the south and the Chesapeake Bay area. Also, the scale is
generally larger for RF-CMAQ equal 50~ppb than 80~ppb. Note that this
does not imply a lighter tail for extreme RF-CMAQ, since the GPD shape
parameter [Figure \ref{fGPD}(b)] is higher for large RF-CMAQ values
compared to moderate RF-CMAQ values.

\begin{figure}

\includegraphics{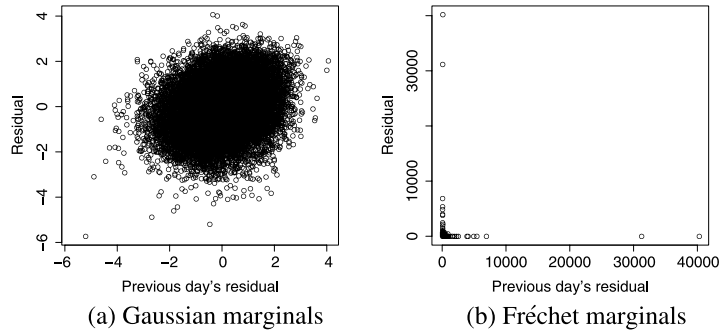}

\caption{Plots of the residuals for consecutive days at the same
locations, transformed to \textup{(a)} standard normal and \textup
{(b)} unit $\Frechet$
marginals.}\label{fresids}
\end{figure}

To determine the form of residual dependence, we compute the estimated
Gaussian-transformed residuals $z(t,\bs)$. Figure \ref{fresids}(a)
plots the residuals for each consecutive pair of observations at the
same station, $[z(t-1,\bs),z(t,\bs)]$. The sample correlation is 0.26
($p$-value for the test of correlation is $<$0.001). We note that much of
the autocorrelation in ozone is captured by the RF-CMAQ model and,
thus, the correlation in the residuals is lower than the correlation in
the raw ozone values. To test for extremal dependence, Figure \ref
{fresids}(b) plots the residuals transformed to have unit $\Frechet$
margins to emphasize dependence in the tails. The unit $\Frechet$
distribution function is $P(Z<c) = \exp(-1/c)$, therefore, if $z(t,\bs
)$ is standard normal, then $z_F(t,\bs) = -1/\log\{\Phi[z(t,\bs)]\}
$ is unit $\Frechet$. The pairs $[z_F(t-1,\bs),z_F(t,\bs)]$ show no
asymptotic dependence since for all pairs with one extremely large
value the other member of the pair is near zero.
Therefore, we use a Gaussian copula for predictive purposes.

\section{The distribution of extremes under various control
strategies}\label{scontrol}

To determine the local effects on extreme ozone events of several
control strategies, we sample $R$ replicates of summer ozone at each
grid cell from the predictive distribution, as described in Section
\ref{spredict}. We compare four control strategies:
\begin{longlist}[S0:]
\item[S0:] the base case with no change in emissions,
$\boldeta
=(0,0,0,0,0,0)$,
\item[S1:] a 50\% reduction in mobile-source NO$_{\mathrm{x}}$,
$\boldeta=(-0.5,0,0,0,0,0)$,
\item[S2:] a 50\% reduction in point-source NO$_{\mathrm{x}}$,
$\boldeta=(0,-0.5,0,0,0,0)$,
\item[S3:] a 15\% reduction in mobile, point, and other-source
NO$_{\mathrm{x}}$, $\boldeta=(-0.15,\break -0.15,-0.15,0,0,0)$.
\end{longlist}
These emission reductions were selected to give roughly a
spatial-average of 3~ppb decrease in the base CMAQ $C_0(\bs,t)$. These
reductions are in line with reductions often considered by regulators
and air quality managers, for example, the 2008 EPA Regulatory Impact
Analysis (\href{http://www.epa.gov/ttn/ecas/regdata/RIAs/452\_R\_08\_003.pdf}{http://www.epa.gov/ttn/}
\href{http://www.epa.gov/ttn/ecas/regdata/RIAs/452\_R\_08\_003.pdf}{ecas/regdata/RIAs/452\_R\_08\_003.pdf}), which considers
reductions of 30\% to 90\% for both VOC and NO$_{\mathrm{x}}$. We display the
predictive distribution by computing various summary statistics for
each replication, for example, $y_4^{(r)}(\bs)$, the fourth largest
value of $\{y(1,\bs),\ldots,y(n_T,\bs)\}$ for replication $r$, and
plotting its mean, $\sum_{r=1}^Ry_4^{(r)}(\bs)/R$, and proportion
above 75~ppb, $\sum_{r=1}^RI[y_4^{(r)}(\bs)>75]/R$.

\begin{figure}

\includegraphics{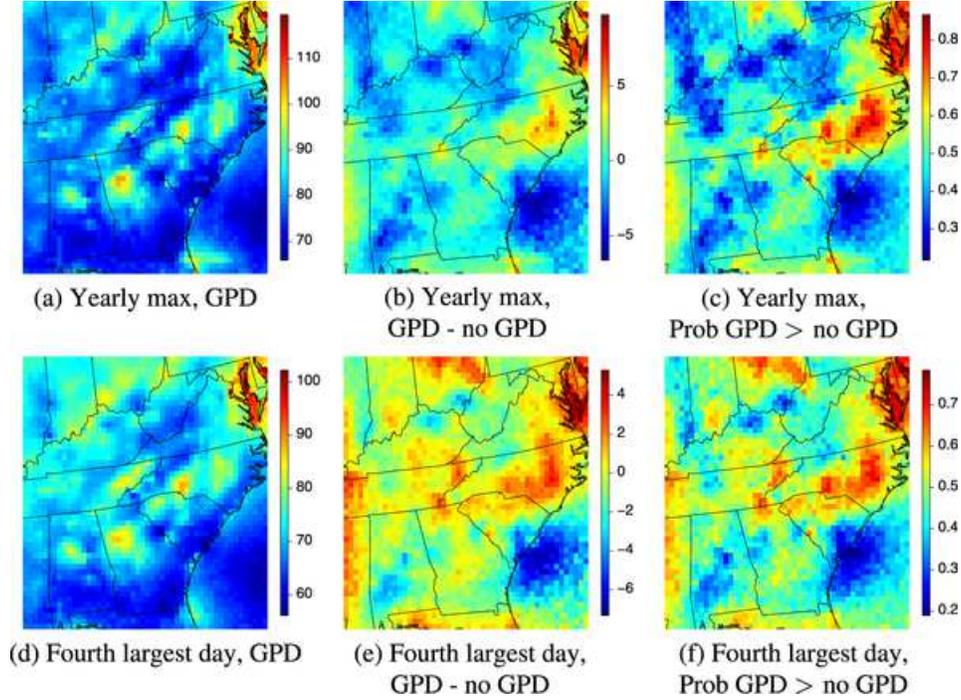}

\caption{Comparison of the predictive distribution for the yearly
maximum and yearly 4th highest value (ppb) for the base case with
(``GPD'') and without (``noGPD'') Generalized Pareto tails. Plotted are
the posterior mean for the GPD model [panels \textup{(a)} and \textup
{(d)}], the
posterior mean difference between GPD and noGPD [panels \textup{(b)}
and \textup{(e)}],
and the posterior probability that GPD gives a larger value than noGPD
[panels \textup{(c)} and \textup{(f)}].}\label{fbasecase}
\end{figure}

To illustrate the effects modeling the tail as GPD rather than
Gaussian, Figure~\ref{fbasecase} plots the average yearly maximum and
fourth highest day for the base case S0 with the final model with
$L=1$, $M=2$, and GPD tails and the fit without the GPD tail, that is,
a Gaussian model with mean $\beta(\mathbf{x},\bs)$ and standard
deviation $\theta_1(\mathbf{x},\bs)$. The two models differ by
3--5~ppb in many locations, which is a meaningful difference for
regulatory purposes.
The probability that the yearly maximum and fourth highest day are
larger using the GPD model is 0.8--0.9 in eastern North Carolina and the
Chesapeake Bay area [Figures \ref{fbasecase}(c) and \ref
{fbasecase}(f)]. Although these probabilities are not definitive, we
note that they are computed using separate samples from the residual
distribution [$\bz(\bs)$ in Section \ref{spredict}] and, therefore,
these probabilities represent an almost complete separation of the
predictive distributions under these two models.

\begin{figure}

\includegraphics{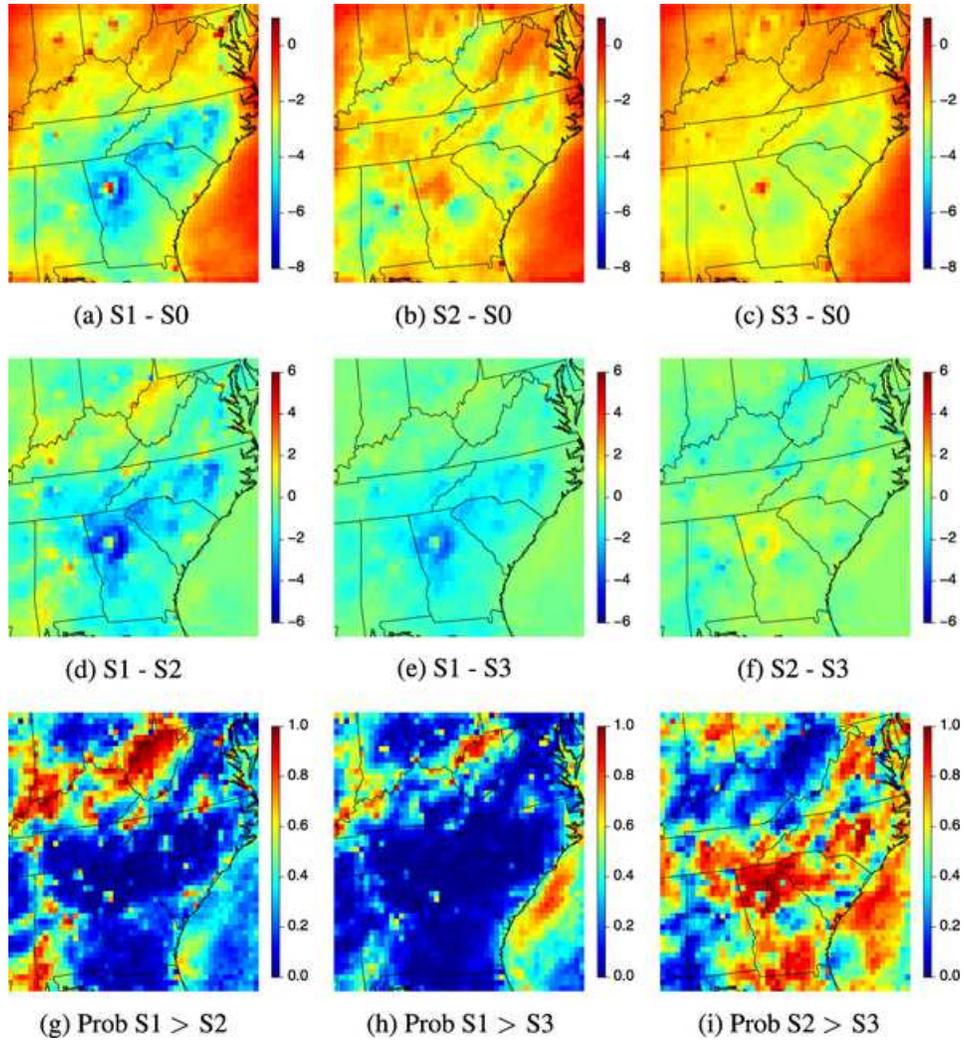}

\caption{Difference in the posterior predictive mean of the fourth
highest day of the year for the several pairs of scenarios [panels
\textup{(a)--(f)}; parts per billion] and probability that the fourth
highest day
of the year is larger for one scenario than another [panels \textup{(g)--(i)}].
The probabilities of a reduction from S0 are near one for all other
scenarios and thus not shown.}\label{fcontrolmeans}
\end{figure}

Figure \ref{fcontrolmeans} compares the projection of the 4th highest
day of the year under the four scenarios. Compared to the base case,
the 50\% reduction in mobile source NO$_{\mathrm{x}}$ has the largest
effects in the area surrounding Atlanta [Figure~\ref
{fcontrolmeans}(a)]. The reduction is as large as 6~ppb to the east of
Atlanta. In the center of the city, however, this control strategy
gives virtually no reduction. It is well known that high NO released in
high-traffic areas destroys ozone near the source and, therefore,
reducing the mobile-source emissions does not reduce ozone near the
source, but rather downwind where NO concentrations are lower. This can
be seen in the map of the sensitivities in Figure \ref{fdata}(c), which
is negative in Atlanta's center, but positive in its southeastern
suburbs. The reduction in ozone for the point-source NO$_{\mathrm{x}}$
scenario is generally smaller and is more uniform across space [Figure
\ref{fcontrolmeans}(b)]. The reduction is 2--5~ppb for most of the region
with exceptions of smaller reductions in Atlanta and Northern Virginia.
Comparing the reductions corresponding to the mobile-source and
point-source control strategies [Figure \ref{fcontrolmeans}(d)], we
find a larger reduction for the mobile-source strategy in most of the
spatial domain, with exceptions in Kentucky, West Virginia, and central
Atlanta. The third control strategy of reducing all the NO$_{\mathrm
{x}}$ emissions by 15\% shows a similar spatial pattern to the
mobile-source strategy, but with generally smaller reductions.

\begin{figure}

\includegraphics{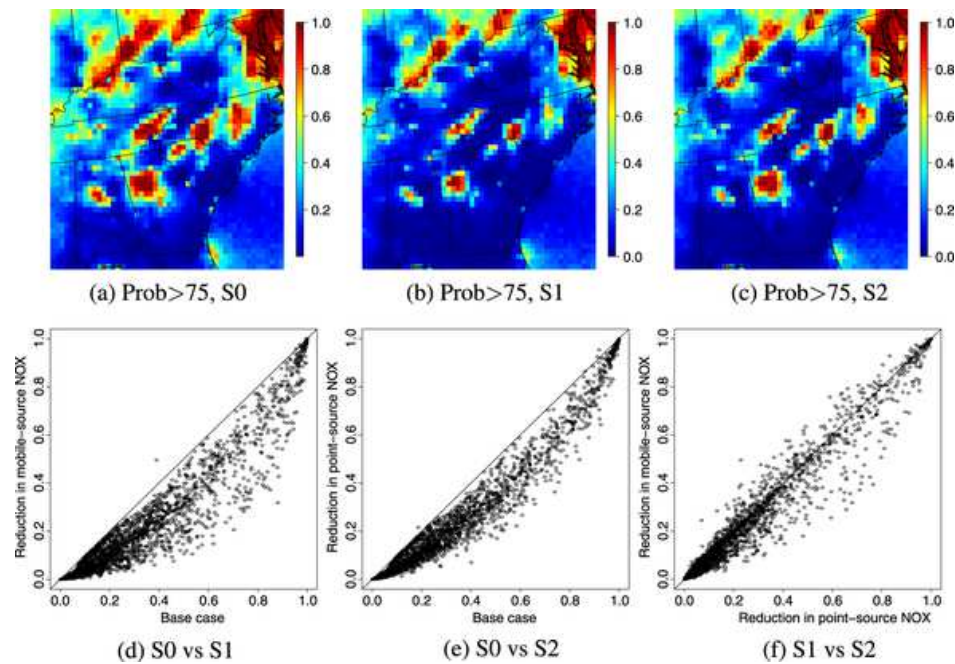}

\caption{Comparison of the predictive probability of greater than
75~ppb for the yearly 4th highest value under three control strategies.
Each point in panels \textup{(d)--(f)} corresponds to one of the grid
cells in
panels \textup{(a)--(c)}.}\label{fcontrolprobs}
\end{figure}

In addition to comparing changes in the mean of the fourth highest day
of the year, the predictive distributions can also be used to study the
probability that the fourth highest day exceeds the current standard of
75~ppb. Figure \ref{fcontrolprobs} shows that many areas have a
substantial reduction in the exceedance probability. For example, the
Birmingham and Raleigh areas go from near 1.0 in the base case to
0.6--0.8 under the mobile-source control strategy. However, areas with
the highest ozone level, Atlanta and Chesapeake Bay, have exceedance
probability near one in all cases. Comparing the mobile-source and
point-source strategies [Figure \ref{fcontrolprobs}(f)], the exceedance
probability is lower under the mobile-source control strategy than the
point-source strategy for 74\% of the grid cells. Overall, the
mobile-sources reduction strategy appears to be the most effective.

\section{Discussion}\label{sconc}

In this paper we propose a new framework for downscaling extremes. We
propose to model the conditional distribution of the monitor data given
the RFM as a combination of quantile regression and extreme value
modeling, using generalized Pareto tails. Using a fully-Bayesian
analysis, we propagate many sources of uncertainty through to the final
estimate of the effect of each control strategy. Using this approach,
we evaluate three control strategies related to reduction in
NO$_{\mathrm
{x}}$. We find that reducing mobile-sources NO$_{\mathrm{x}}$ has the
largest impact of the strategies considered, especially in suburban
Atlanta. However, the probability of noncompliance with the EPA
regulation remains near one for the Atlanta area for all control strategies.

Although our modeling framework is quite flexible, it has several
limitations. First we have not considered residual spatial correlation
in the model fitting stage. \citet{Reich-2012} does include residual
spatiotemporal dependence for spatial quantile regression. Although
estimating residual spatial dependence in not the primary focus in this
work, failing to account for this may cause underestimation of the
uncertainty of model parameters. Also, although the quantile function
is modeled as increasing for each value of the RFM, the quantile
function is not forced to be increasing in RFM for each quantile level,
as one would naturally expect. Including this prior belief may be
possible by adding further restrictions to the polynomial coefficients
in the conditional distribution and would certainly improve the fit for
small and moderate data sets.

Also, we note that our analysis only considers global emission control
strategies that assume a uniform reduction in emissions across the
entire spatial domain. An extension would be to calculate sensitivities
to local changes, to study the effects of emission reduction in one
grid cell on neighboring cells. This would add another spatial aspect
to the model and provide a more comprehensive sensitivity analysis.

\section*{Acknowledgments}

The authors wish to thank the Editor, Associate Editor, and anonymous
referees for their thoughtful comments which greatly improved the
manuscript.

%



\printaddresses


\begin{thebibliography}{36}

\bibitem[\protect\citeauthoryear{Behrens, Lopes and
  Gamerman}{2004}]{behrens-2004a}
\begin{barticle}[mr]
\bauthor{\bsnm{Behrens},~\bfnm{Cibele~N.}\binits{C.~N.}},
  \bauthor{\bsnm{Lopes},~\bfnm{Hedibert~F.}\binits{H.~F.}} \AND
  \bauthor{\bsnm{Gamerman},~\bfnm{Dani}\binits{D.}}
(\byear{2004}).
\btitle{Bayesian analysis of extreme events with threshold estimation}.
\bjournal{Stat. Model.}
\bvolume{4}
\bpages{227--244}.
\bid{doi={10.1191/1471082X04st075oa}, issn={1471-082X}, mr={2062102}}
\bptok{imsref}%
\end{barticle}
\endbibitem\eject

\bibitem[\protect\citeauthoryear{Bentzien and
  Friederichs}{2012}]{Bentzien-2012}
\begin{barticle}[auto:STB|2013/04/11|08:11:48]
\bauthor{\bsnm{Bentzien},~\bfnm{S.}\binits{S.}} \AND
  \bauthor{\bsnm{Friederichs},~\bfnm{P.}\binits{P.}}
(\byear{2012}).
\btitle{Generating and calibrating probabilistic quantitative precipitation
  forecasts from the high-resolution nwp model cosmo-de}.
\bjournal{Weather and Forecasting}
\bvolume{27}
\bpages{998--1002}.
\bptok{imsref}%
\end{barticle}
\endbibitem

\bibitem[\protect\citeauthoryear{Berrocal, Gelfand and
  Holland}{2010}]{berrocal-2010a}
\begin{barticle}[mr]
\bauthor{\bsnm{Berrocal},~\bfnm{Veronica~J.}\binits{V.~J.}},
  \bauthor{\bsnm{Gelfand},~\bfnm{Alan~E.}\binits{A.~E.}} \AND
  \bauthor{\bsnm{Holland},~\bfnm{David~M.}\binits{D.~M.}}
(\byear{2010}).
\btitle{A spatio-temporal downscaler for output from numerical models}.
\bjournal{J. Agric. Biol. Environ. Stat.}
\bvolume{15}
\bpages{176--197}.
\bid{doi={10.1007/s13253-009-0004-z}, issn={1085-7117}, mr={2787270}}
\bptok{imsref}%
\end{barticle}
\endbibitem

\bibitem[\protect\citeauthoryear{Byun and Schere}{2006}]{Byun2006}
\begin{barticle}[auto:STB|2013/04/11|08:11:48]
\bauthor{\bsnm{Byun},~\bfnm{D.}\binits{D.}} \AND
  \bauthor{\bsnm{Schere},~\bfnm{K.~L.}\binits{K.~L.}}
(\byear{2006}).
\btitle{Review of the governing equations, computational algorithms, and other
  components of the models-3 community multiscale air quality (cmaq) modeling
  system}.
\bjournal{Applied Mechanics Reviews}
\bvolume{59}
\bpages{51--77}.
\bptok{imsref}%
\end{barticle}
\endbibitem

\bibitem[\protect\citeauthoryear{Carter}{2000}]{Carter2000}
\begin{bmisc}[auto:STB|2013/04/11|08:11:48]
\bauthor{\bsnm{Carter},~\bfnm{W.~P.~L.}\binits{W.~P.~L.}}
(\byear{2000}).
\bhowpublished{Implementation of the {SAPRC-99} chemical mechanism into the
  models-3 framework. Report to the United States Environmental Protection
  Agency. Available at \url{http://www.cert.ucr.edu/\textasciitilde
  carter/pubs/s99mod3.pdf}.}
\bptok{imsref}%
\end{bmisc}
\endbibitem

\bibitem[\protect\citeauthoryear{Chavez-Demoulin and
  Davison}{2012}]{chavez-2012}
\begin{barticle}[mr]
\bauthor{\bsnm{Chavez-Demoulin},~\bfnm{V.}\binits{V.}} \AND
  \bauthor{\bsnm{Davison},~\bfnm{A.~C.}\binits{A.~C.}}
(\byear{2012}).
\btitle{Modelling time series extremes}.
\bjournal{REVSTAT}
\bvolume{10}
\bpages{109--133}.
\bid{issn={1645-6726}, mr={2912373}}
\bptok{imsref}%
\end{barticle}
\endbibitem

\bibitem[\protect\citeauthoryear{Cohan et~al.}{2005}]{Cohan2005}
\begin{barticle}[auto:STB|2013/04/11|08:11:48]
\bauthor{\bsnm{Cohan},~\bfnm{D.~S.}\binits{D.~S.}},
  \bauthor{\bsnm{Hakami},~\bfnm{A.}\binits{A.}},
  \bauthor{\bsnm{Hu},~\bfnm{Y.~T.}\binits{Y.~T.}} \AND
  \bauthor{\bsnm{Russell},~\bfnm{A.~G.}\binits{A.~G.}}
(\byear{2005}).
\btitle{Nonlinear response of ozone to emissions: Source apportionment and
  sensitivity analysis}.
\bjournal{Environmental Science and Technology}
\bvolume{39}
\bpages{6739--6748}.
\bptok{imsref}%
\end{barticle}
\endbibitem

\bibitem[\protect\citeauthoryear{Coles}{2001}]{coles-2001a}
\begin{bbook}[mr]
\bauthor{\bsnm{Coles},~\bfnm{Stuart}\binits{S.}}
(\byear{2001}).
\btitle{An Introduction to Statistical Modeling of Extreme Values}.
\bpublisher{Springer}, \blocation{London}.
\bid{mr={1932132}}
\bptok{imsref}%
\end{bbook}
\endbibitem

\bibitem[\protect\citeauthoryear{Cooley and Sain}{2010}]{cooley-2010a}
\begin{barticle}[mr]
\bauthor{\bsnm{Cooley},~\bfnm{Daniel}\binits{D.}} \AND
  \bauthor{\bsnm{Sain},~\bfnm{Stephan~R.}\binits{S.~R.}}
(\byear{2010}).
\btitle{Spatial hierarchical modeling of precipitation extremes from a regional
  climate model}.
\bjournal{J. Agric. Biol. Environ. Stat.}
\bvolume{15}
\bpages{381--402}.
\bid{doi={10.1007/s13253-010-0023-9}, issn={1085-7117}, mr={2787265}}
\bptok{imsref}%
\end{barticle}
\endbibitem

\bibitem[\protect\citeauthoryear{Davison and Smith}{1990}]{davison-1990}
\begin{barticle}[mr]
\bauthor{\bsnm{Davison},~\bfnm{A.~C.}\binits{A.~C.}} \AND
  \bauthor{\bsnm{Smith},~\bfnm{R.~L.}\binits{R.~L.}}
(\byear{1990}).
\btitle{Models for exceedances over high thresholds}.
\bjournal{J. R. Stat. Soc. Ser. B Stat. Methodol.}
\bvolume{52}
\bpages{393--442}.
\bid{issn={0035-9246}, mr={1086795}}
\bptnote{check related}%
\bptok{imsref}%
\end{barticle}
\endbibitem

\bibitem[\protect\citeauthoryear{Digar et~al.}{2011}]{digar-2011}
\begin{barticle}[auto:STB|2013/04/11|08:11:48]
\bauthor{\bsnm{Digar},~\bfnm{A.}\binits{A.}},
  \bauthor{\bsnm{Cohan},~\bfnm{D.}\binits{D.}},
  \bauthor{\bsnm{Cox},~\bfnm{D.}\binits{D.}},
  \bauthor{\bsnm{Byeong-Uk},~\bfnm{K.}\binits{K.}} \AND
  \bauthor{\bsnm{Boylan},~\bfnm{J.}\binits{J.}}
(\byear{2011}).
\btitle{Likelihood of achieving air quality targets under model uncertainties}.
\bjournal{Environ. Sci. Technol.}
\bvolume{45}
\bpages{189--196}.
\bptok{imsref}%
\end{barticle}
\endbibitem

\bibitem[\protect\citeauthoryear{Eastoe and Tawn}{2012}]{eastoe-2012}
\begin{barticle}[mr]
\bauthor{\bsnm{Eastoe},~\bfnm{Emma~F.}\binits{E.~F.}} \AND
  \bauthor{\bsnm{Tawn},~\bfnm{Jonathan~A.}\binits{J.~A.}}
(\byear{2012}).
\btitle{Modelling the distribution of the cluster maxima of exceedances of
  subasymptotic thresholds}.
\bjournal{Biometrika}
\bvolume{99}
\bpages{43--55}.
\bid{doi={10.1093/biomet/asr078}, issn={0006-3444}, mr={2899662}}
\bptok{imsref}%
\end{barticle}
\endbibitem

\bibitem[\protect\citeauthoryear{Foley, Reich and Napelenok}{2012}]{foley-2012}
\begin{barticle}[auto:STB|2013/04/11|08:11:48]
\bauthor{\bsnm{Foley},~\bfnm{K.~M.}\binits{K.~M.}},
  \bauthor{\bsnm{Reich},~\bfnm{B.~J.}\binits{B.~J.}} \AND
  \bauthor{\bsnm{Napelenok},~\bfnm{S.~L.}\binits{S.~L.}}
(\byear{2012}).
\btitle{Bayesian analysis of a reduced-form air quality model}.
\bjournal{Environmental Science \& Technology}
\bvolume{46}
\bpages{7604--7611}.
\bptok{imsref}%
\end{barticle}
\endbibitem

\bibitem[\protect\citeauthoryear{Foley et~al.}{2010}]{Foley2010}
\begin{barticle}[auto:STB|2013/04/11|08:11:48]
\bauthor{\bsnm{Foley},~\bfnm{K.~M.}\binits{K.~M.}},
  \bauthor{\bsnm{Roselle},~\bfnm{S.~J.}\binits{S.~J.}},
  \bauthor{\bsnm{Appel},~\bfnm{K.~W.}\binits{K.~W.}},
  \bauthor{\bsnm{Bhave},~\bfnm{P.~V.}\binits{P.~V.}},
  \bauthor{\bsnm{Pleim},~\bfnm{J.~E.}\binits{J.~E.}},
  \bauthor{\bsnm{Otte},~\bfnm{T.~L.}\binits{T.~L.}},
  \bauthor{\bsnm{Mathur},~\bfnm{R.}\binits{R.}},
  \bauthor{\bsnm{Sarwar},~\bfnm{G.}\binits{G.}},
  \bauthor{\bsnm{Young},~\bfnm{J.~O.}\binits{J.~O.}},
  \bauthor{\bsnm{Gilliam},~\bfnm{R.~C.}\binits{R.~C.}},
  \bauthor{\bsnm{Nolte},~\bfnm{C.~G.}\binits{C.~G.}},
  \bauthor{\bsnm{Kelly},~\bfnm{J.~T.}\binits{J.~T.}},
  \bauthor{\bsnm{Gilliland},~\bfnm{A.~B.}\binits{A.~B.}} \AND
  \bauthor{\bsnm{Bash},~\bfnm{J.~O.}\binits{J.~O.}}
(\byear{2010}).
\btitle{Incremental testing of the community multiscale air quality {(CMAQ)}
  modeling system version 4.7}.
\bjournal{Geoscientific Model Development}
\bvolume{3}
\bpages{204--226}.
\bptok{imsref}%
\end{barticle}
\endbibitem

\bibitem[\protect\citeauthoryear{Frigessi, Haug and Rue}{2002}]{Frigessi-2003}
\begin{barticle}[mr]
\bauthor{\bsnm{Frigessi},~\bfnm{Arnoldo}\binits{A.}},
  \bauthor{\bsnm{Haug},~\bfnm{Ola}\binits{O.}} \AND
  \bauthor{\bsnm{Rue},~\bfnm{H{\aa}vard}\binits{H.}}
(\byear{2002}).
\btitle{A dynamic mixture model for unsupervised tail estimation without
  threshold selection}.
\bjournal{Extremes}
\bvolume{5}
\bpages{219--235}.
\bid{doi={10.1023/A:1024072610684}, issn={1386-1999}, mr={1995776}}
\bptnote{check year}%
\bptok{imsref}%
\end{barticle}
\endbibitem

\bibitem[\protect\citeauthoryear{Fuentes, Henry and
  Reich}{2013}]{fuentes-2010a}
\begin{barticle}[mr]
\bauthor{\bsnm{Fuentes},~\bfnm{Montserrat}\binits{M.}},
  \bauthor{\bsnm{Henry},~\bfnm{John}\binits{J.}} \AND
  \bauthor{\bsnm{Reich},~\bfnm{Brian}\binits{B.}}
(\byear{2013}).
\btitle{Nonparametric spatial models for extremes: Application to extreme
  temperature data}.
\bjournal{Extremes}
\bvolume{16}
\bpages{75--101}.
\bid{doi={10.1007/s10687-012-0154-1}, issn={1386-1999}, mr={3020178}}
\bptnote{check year}%
\bptok{imsref}%
\end{barticle}
\endbibitem

\bibitem[\protect\citeauthoryear{Gneiting and Raftery}{2007}]{Gneiting-2007}
\begin{barticle}[mr]
\bauthor{\bsnm{Gneiting},~\bfnm{Tilmann}\binits{T.}} \AND
  \bauthor{\bsnm{Raftery},~\bfnm{Adrian~E.}\binits{A.~E.}}
(\byear{2007}).
\btitle{Strictly proper scoring rules, prediction, and estimation}.
\bjournal{J. Amer. Statist. Assoc.}
\bvolume{102}
\bpages{359--378}.
\bid{doi={10.1198/016214506000001437}, issn={0162-1459}, mr={2345548}}
\bptok{imsref}%
\end{barticle}
\endbibitem

\bibitem[\protect\citeauthoryear{Grell, Dudhia and Stauffer}{1994}]{Grell1995}
\begin{bmisc}[auto:STB|2013/04/11|08:11:48]
\bauthor{\bsnm{Grell},~\bfnm{G.~A.}\binits{G.~A.}},
  \bauthor{\bsnm{Dudhia},~\bfnm{A.~J.}\binits{A.~J.}} \AND
  \bauthor{\bsnm{Stauffer},~\bfnm{D.~R.}\binits{D.~R.}}
(\byear{1994}).
\bhowpublished{A description of the fifth-generation pennstate/ncar mesoscale
  model (mm5). NCAR Technical Note NCAR/TN-398+STR. Available at
  \url{http://www.mmm.ucar.edu/mm5/doc1.html}.}
\bptok{imsref}%
\end{bmisc}
\endbibitem

\bibitem[\protect\citeauthoryear{Higdon et~al.}{2004}]{Higdon-2004}
\begin{barticle}[mr]
\bauthor{\bsnm{Higdon},~\bfnm{Dave}\binits{D.}},
  \bauthor{\bsnm{Kennedy},~\bfnm{Marc}\binits{M.}},
  \bauthor{\bsnm{Cavendish},~\bfnm{James~C.}\binits{J.~C.}},
  \bauthor{\bsnm{Cafeo},~\bfnm{John~A.}\binits{J.~A.}} \AND
  \bauthor{\bsnm{Ryne},~\bfnm{Robert~D.}\binits{R.~D.}}
(\byear{2004}).
\btitle{Combining field data and computer simulations for calibration and
  prediction}.
\bjournal{SIAM J. Sci. Comput.}
\bvolume{26}
\bpages{448--466}.
\bid{doi={10.1137/S1064827503426693}, issn={1064-8275}, mr={2116355}}
\bptok{imsref}%
\end{barticle}
\endbibitem

\bibitem[\protect\citeauthoryear{Kennedy and O'Hagan}{2001}]{Kennedy-2001}
\begin{barticle}[mr]
\bauthor{\bsnm{Kennedy},~\bfnm{Marc~C.}\binits{M.~C.}} \AND
  \bauthor{\bsnm{O'Hagan},~\bfnm{Anthony}\binits{A.}}
(\byear{2001}).
\btitle{Bayesian calibration of computer models}.
\bjournal{J.~R.~Stat. Soc. Ser. B Stat. Methodol.}
\bvolume{63}
\bpages{425--464}.
\bid{doi={10.1111/1467-9868.00294}, issn={1369-7412}, mr={1858398}}
\bptnote{check related}%
\bptok{imsref}%
\end{barticle}
\endbibitem

\bibitem[\protect\citeauthoryear{Kharin et~al.}{2007}]{kharin-2007a}
\begin{barticle}[auto:STB|2013/04/11|08:11:48]
\bauthor{\bsnm{Kharin},~\bfnm{V.}\binits{V.}},
  \bauthor{\bsnm{Zwiers},~\bfnm{F.}\binits{F.}},
  \bauthor{\bsnm{Zhang},~\bfnm{X.}\binits{X.}} \AND
  \bauthor{\bsnm{Hegerl},~\bfnm{G.}\binits{G.}}
(\byear{2007}).
\btitle{Changes in temperature and precipitation extremes in the IPCC ensemble
  of global coupled model simulations}.
\bjournal{J. Climate}
\bvolume{20}
\bpages{1419--1444}.
\bptok{imsref}%
\end{barticle}
\endbibitem

\bibitem[\protect\citeauthoryear{Mannshardt-Shamseldin
  et~al.}{2010}]{mannshardt-2010a}
\begin{barticle}[mr]
\bauthor{\bsnm{Mannshardt-Shamseldin},~\bfnm{Elizabeth~C.}\binits{E.~C.}},
  \bauthor{\bsnm{Smith},~\bfnm{Richard~L.}\binits{R.~L.}},
  \bauthor{\bsnm{Sain},~\bfnm{Stephan~R.}\binits{S.~R.}},
  \bauthor{\bsnm{Mearns},~\bfnm{Linda~O.}\binits{L.~O.}} \AND
  \bauthor{\bsnm{Cooley},~\bfnm{Daniel}\binits{D.}}
(\byear{2010}).
\btitle{Downscaling extremes: A comparison of extreme value distributions in
  point-source and gridded precipitation data}.
\bjournal{Ann. Appl. Stat.}
\bvolume{4}
\bpages{484--502}.
\bid{doi={10.1214/09-AOAS287}, issn={1932-6157}, mr={2758181}}
\bptok{imsref}%
\end{barticle}
\endbibitem

\bibitem[\protect\citeauthoryear{Maraun, Osborn and Rust}{2011}]{maraun-2011a}
\begin{barticle}[auto:STB|2013/04/11|08:11:48]
\bauthor{\bsnm{Maraun},~\bfnm{D.}\binits{D.}},
  \bauthor{\bsnm{Osborn},~\bfnm{T.}\binits{T.}} \AND
  \bauthor{\bsnm{Rust},~\bfnm{H.}\binits{H.}}
(\byear{2011}).
\btitle{The influence of synoptic airflow on UK daily precipitation extremes.
  Part I: Observed spatio-temporal relationships}.
\bjournal{Climate Dynamics}
\bvolume{36}
\bpages{261--275}.
\bptok{imsref}%
\end{barticle}
\endbibitem

\bibitem[\protect\citeauthoryear{Napelenok et~al.}{2011}]{Napelenok-2011}
\begin{barticle}[auto:STB|2013/04/11|08:11:48]
\bauthor{\bsnm{Napelenok},~\bfnm{S.}\binits{S.}},
  \bauthor{\bsnm{Foley},~\bfnm{K.}\binits{K.}},
  \bauthor{\bsnm{Kang},~\bfnm{D.}\binits{D.}},
  \bauthor{\bsnm{Mathur},~\bfnm{R.}\binits{R.}},
  \bauthor{\bsnm{Pierce},~\bfnm{T.}\binits{T.}} \AND
  \bauthor{\bsnm{Rao},~\bfnm{S.}\binits{S.}}
(\byear{2011}).
\btitle{Dynamic evaluation of regional air quality model's response to
  emissions reduction in the presence of uncertain emission inventories}.
\bjournal{Atmospheric Environment}
\bvolume{45}
\bpages{4091--4098}.
\bptok{imsref}%
\end{barticle}
\endbibitem

\bibitem[\protect\citeauthoryear{Nelsen}{1999}]{Nelson99}
\begin{bbook}[mr]
\bauthor{\bsnm{Nelsen},~\bfnm{Roger~B.}\binits{R.~B.}}
(\byear{1999}).
\btitle{An Introduction to Copulas}.
\bseries{Lecture Notes in Statistics}
\bvolume{139}.
\bpublisher{Springer}, \blocation{New York}.
\bid{mr={1653203}}
\bptok{imsref}%
\end{bbook}
\endbibitem

\bibitem[\protect\citeauthoryear{Reich}{2012}]{Reich-2012}
\begin{barticle}[mr]
\bauthor{\bsnm{Reich},~\bfnm{Brian~J.}\binits{B.~J.}}
(\byear{2012}).
\btitle{Spatiotemporal quantile regression for detecting distributional changes
  in environmental processes}.
\bjournal{J.~R. Stat. Soc. Ser. C. Appl. Stat.}
\bvolume{61}
\bpages{535--553}.
\bid{doi={10.1111/j.1467-9876.2011.01025.x}, issn={0035-9254}, mr={2960737}}
\bptok{imsref}%
\end{barticle}
\endbibitem

\bibitem[\protect\citeauthoryear{Reich, Fuentes and Dunson}{2011}]{reich-2011}
\begin{barticle}[mr]
\bauthor{\bsnm{Reich},~\bfnm{Brian~J.}\binits{B.~J.}},
  \bauthor{\bsnm{Fuentes},~\bfnm{Montserrat}\binits{M.}} \AND
  \bauthor{\bsnm{Dunson},~\bfnm{David~B.}\binits{D.~B.}}
(\byear{2011}).
\btitle{Bayesian spatial quantile regression}.
\bjournal{J.~Amer. Statist. Assoc.}
\bvolume{106}
\bpages{6--20}.
\bid{doi={10.1198/jasa.2010.ap09237}, issn={0162-1459}, mr={2816698}}
\bptok{imsref}%
\end{barticle}
\endbibitem

\bibitem[\protect\citeauthoryear{Schliep et~al.}{2010}]{schliep-2010a}
\begin{barticle}[mr]
\bauthor{\bsnm{Schliep},~\bfnm{Erin~M.}\binits{E.~M.}},
  \bauthor{\bsnm{Cooley},~\bfnm{Daniel}\binits{D.}},
  \bauthor{\bsnm{Sain},~\bfnm{Stephan~R.}\binits{S.~R.}} \AND
  \bauthor{\bsnm{Hoeting},~\bfnm{Jennifer~A.}\binits{J.~A.}}
(\byear{2010}).
\btitle{A comparison study of extreme precipitation from six different regional
  climate models via spatial hierarchical modeling}.
\bjournal{Extremes}
\bvolume{13}
\bpages{219--239}.
\bid{doi={10.1007/s10687-009-0098-2}, issn={1386-1999}, mr={2643558}}
\bptok{imsref}%
\end{barticle}
\endbibitem

\bibitem[\protect\citeauthoryear{Schwede, Pouliot and
  Pierce}{2005}]{Schwede2005}
\begin{bmisc}[auto:STB|2013/04/11|08:11:48]
\bauthor{\bsnm{Schwede},~\bfnm{D.}\binits{D.}},
  \bauthor{\bsnm{Pouliot},~\bfnm{G.~A.}\binits{G.~A.}} \AND
  \bauthor{\bsnm{Pierce},~\bfnm{T.}\binits{T.}}
(\byear{2005}).
\bhowpublished{Changes to the biogenic emissions inventory system version 3
  (BEIS3). In \textit{Proceedings of the 4th CMAS Models-3 Users' Conference},
  26--28 September. Chapel Hill, NC}.
\bptok{imsref}%
\end{bmisc}
\endbibitem

\bibitem[\protect\citeauthoryear{Seinfeld and Pandis}{1998}]{Seinfeld98}
\begin{bbook}[auto:STB|2013/04/11|08:11:48]
\bauthor{\bsnm{Seinfeld},~\bfnm{J.}\binits{J.}} \AND
  \bauthor{\bsnm{Pandis},~\bfnm{S.}\binits{S.}}
(\byear{1998}).
\btitle{Atmospheric Chemistry and Physics}.
\bpublisher{Wiley}, \blocation{New York}.
\bptok{imsref}%
\end{bbook}
\endbibitem

\bibitem[\protect\citeauthoryear{Sillman et~al.}{2011}]{sillmann-2011a}
\begin{barticle}[auto:STB|2013/04/11|08:11:48]
\bauthor{\bsnm{Sillman},~\bfnm{J.}\binits{J.}},
  \bauthor{\bsnm{Croci-Maspoli},~\bfnm{M.}\binits{M.}},
  \bauthor{\bsnm{Kallache},~\bfnm{M.}\binits{M.}} \AND
  \bauthor{\bsnm{Katz},~\bfnm{R.~W.}\binits{R.~W.}}
(\byear{2011}).
\btitle{Extreme cold winter temperatures in Europe under the influence of North
  Atlantic atmospheric blocking}.
\bjournal{J. Climate}
\bvolume{24}
\bpages{5899--5913}.
\bptok{imsref}%
\end{barticle}
\endbibitem

\bibitem[\protect\citeauthoryear{U.S. Environmental Protection~Agency (US
  EPA)}{2006}]{epa-2006}
\begin{bmisc}[auto:STB|2013/04/11|08:11:48]
\borganization{U.S. Environmental Protection Agency (US EPA)}
(\byear{2006}).
\bhowpublished{Air quality criteria for ozone and related photochemical
  oxidants (Final). Washington, DC: U.S. EPA. Available at
  \url{http://cfpub.epa.gov/ncea/CFM/recordisplay.cfm?deid=149923}.}
\bptok{imsref}%
\end{bmisc}
\endbibitem

\bibitem[\protect\citeauthoryear{Wehner}{2005}]{wehner-2005a}
\begin{barticle}[auto:STB|2013/04/11|08:11:48]
\bauthor{\bsnm{Wehner},~\bfnm{M.}\binits{M.}}
(\byear{2005}).
\btitle{Changes in daily precipitation and surface air temperature extremes in
  the {IPCC AR4} models}.
\bjournal{US CLIVAR Variations}
\bvolume{3}
\bpages{5--9}.
\bptok{imsref}%
\end{barticle}
\endbibitem

\bibitem[\protect\citeauthoryear{Wehner et~al.}{2010}]{wehner-2010a}
\begin{barticle}[auto:STB|2013/04/11|08:11:48]
\bauthor{\bsnm{Wehner},~\bfnm{M.}\binits{M.}},
  \bauthor{\bsnm{Smith},~\bfnm{R.}\binits{R.}},
  \bauthor{\bsnm{Bala},~\bfnm{G.}\binits{G.}} \AND
  \bauthor{\bsnm{Duffy},~\bfnm{P.}\binits{P.}}
(\byear{2010}).
\btitle{The effect of horizontal resolution on simulation of very extreme US
  precipitation events in a global atmosphere model}.
\bjournal{Climate Dynamics}
\bvolume{34}
\bpages{241--247}.
\bptok{imsref}%
\end{barticle}
\endbibitem

\bibitem[\protect\citeauthoryear{Zhou, Fuentes and Davis}{2011}]{zhou-2011a}
\begin{barticle}[mr]
\bauthor{\bsnm{Zhou},~\bfnm{Jingwen}\binits{J.}},
  \bauthor{\bsnm{Fuentes},~\bfnm{Montserrat}\binits{M.}} \AND
  \bauthor{\bsnm{Davis},~\bfnm{Jerry}\binits{J.}}
(\byear{2011}).
\btitle{Calibration of numerical model output using nonparametric spatial
  density functions}.
\bjournal{J. Agric. Biol. Environ. Stat.}
\bvolume{16}
\bpages{531--553}.
\bid{doi={10.1007/s13253-011-0076-4}, issn={1085-7117}, mr={2862297}}
\bptnote{check year}%
\bptok{imsref}%
\end{barticle}
\endbibitem

\end{thebibliography}
\end{document}